\documentstyle[12pt]{article}
\def\lsim{\mathrel{\rlap {\raise.5ex\hbox{$ < $}}
{\lower.5ex\hbox{$\sim$}}}}

\newcommand{\pr}{\paragraph{}}
\newcommand{\be}{\begin{equation}}
\newcommand{\ee}{\end{equation}}
\newcommand{\bea}{\begin{eqnarray}}
\newcommand{\nn}{\nonumber}
\newcommand{\eea}{\end{eqnarray}}
\newcommand{\nd}[1]{/\hspace{-0.6em} #1}
\newcommand{\nk}{\noindent}
\def\gappeq{\mathrel{\rlap {\raise.5ex\hbox{$>$}}
{\lower.5ex\hbox{$\sim$}}}}
 
\def\lappeq{\mathrel{\rlap{\raise.5ex\hbox{$<$}}
{\lower.5ex\hbox{$\sim$}}}}
 
\begin{document}
 
\begin{titlepage}

\begin{flushright}
ACT-04/96 \\
CERN-TH/96-81 \\
CTP-TAMU-11/96 \\
OUTP-96-15P \\
hep-th/9605046 \\
\end{flushright}
\begin{centering}
\vspace{.1in}
{\large {\bf $D$ Branes from Liouville Strings
}} \\
\vspace{.2in}
{\bf John Ellis$^{a}$},
{\bf N.E. Mavromatos$^{b}$}
 and
{\bf D.V. Nanopoulos$^{c}$}
\\
\vspace{0.2in}
\nk $^a$ Theory Division, CERN, CH-1211, Geneva, Switzerland,  \\
$^b$ University of Oxford, Dept. of Physics
(Theoretical Physics),
1 Keble Road, Oxford OX1 3NP, United Kingdom (P.P.A.R.C. Advanced
Fellow),   \\
$^{c}$ Center for
Theoretical Physics, Dept. of Physics,
Texas A \& M University, College Station, TX 77843-4242, USA
and Astroparticle Physics Group, Houston
Advanced Research Center (HARC), The Mitchell Campus,
Woodlands, TX 77381, USA. \\

\vspace{.03in}
\vspace{.1in}
{\bf Abstract} \\
\vspace{.05in}
\end{centering}
{\small  We develop quantization aspects of our Liouville
approach to non-critical strings, proposing a path-integral
formulation of a   second quantization      of string theory,
that incorporates  naturally the couplings of
string sources to background fields.
Such couplings are characteristic of
macroscopic string solutions and/or
$D$-brane theories.
Resummation over world-sheet genera in the presence of
stringy ($\sigma$-model) soliton backgrounds,
and recoil effects associated with logarithmic operators on the
world sheet,
play a crucial r\^ole in inducing
such sources as well-defined renormalization-group
counterterms.
Using our Liouville renormalization
group approach, we derive the appropriate
second-order equation of motion for the $D$ brane.
We discuss within this approach
the appearance of open strings, whose ends carry
non-trivial Chan-Paton-like quantum numbers
related to the $W_\infty$ charges of two-dimensional
string black holes.}
\vspace{0.4in}
\begin{flushleft}
ACT-04/96 \\
CERN-TH/96-81 \\
CTP-TAMU-11/96 \\
OUTP-96-15P \\
May 1996 \\
\end{flushleft}
\end{titlepage}
\newpage
\section{Introduction}
\pr
    A key issue in string theory is how to go beyond perturbation
theory in a fixed space-time background, describable as one of
many apparently consistent  classical
string vacua,  each of which is characterized by a unique
conformal field theory on the world sheet. One would like to
understand how to interpolate between such backgrounds, and
how to treat the non-perturbative transitions between them. A
related issue is how to treat fluctuations in the topology of
the string world sheet, which appear in the integration over
higher genera.
\pr
    Resolution of these two issues necessarily involves a
departure from criticality. This is because, on the one hand,
 transitions between different
classical vacua, i.e., critical string theories, must traverse
some broader framework in which these are embedded. Moreover, as
we discuss in more detail later,
quantum fluctuations in higher genera may in general be viewed
as sources that
perturb the world-sheet field theory away from criticality.
The minimal broader framework that can be considered is that
of general renormalizable
two-dimensional theories on the world sheet.
        Renormalization      may be implemented using the
Liouville field on the world sheet as a local renormalization
scale~\cite{emn}. A suitable version of the Zamolodchikov
metric~\cite{zam}
can be
introduced as a positive metric on the relevant
space of unitary two-dimensional field theories, and their
renormalization group flow is monotonic with respect to a
suitable version of the Zamolodchikov $C$ function~\cite{zam}.
 
\pr
We have shown~\cite{emn} how the Liouville field may be identified
with the time variable~\cite{emn,aben,dvn}.
Starting from a generalized light-cone
formulation of critical string theory, in which time does not
appear as a target coordinate, we have shown that
renormalization counterterms associated with transitions to
supercritical string theories~\cite{aben}
naturally yield a negative
metric for the Liouville field. An explicit calculation within
this framework reproduces the metric of the
two-dimensional (spherically-symmetric four-dimensional)
string black hole~\cite{witt}.
The monotonicity of the Zamolodchikov $C$ function
and the semigroup nature of the renormalization group
translate
into a monotonic increase in the entropy of the effective
target-space theory~\cite{emn,kogan}.
 
\pr
The Zamolodchikov renormalization-group flow through the
space of world-sheet theories is itself subject to quantum
fluctuations, induced in particular by the sum over higher
genera mentioned above. We have demonstrated that
this flow satisfies the
Helmholtz conditions for consistent first
quantization~\cite{emninfl}.
This result suggests that it may be possible to
define suitably a
functional integral over configurations of the world sheet.
The purpose of this paper is to embark on this programme
and to derive as a first consequence the action of a $D$ brane
coupled as a source to the string effective action~\cite{gibbons}~(in standard notation):
\bea
&~& S=\int d^Dy \sqrt{g} [R + 4 (\partial \Phi )^2 - \frac{1}{12}
e^{-2\Phi} H_{MNP}^2 + \dots ]  + \nn \\
&~&\frac{1}{4\pi \alpha '}\int_\Sigma  d^2\sigma [
\partial_\alpha X^M \partial^\alpha X^N e^{\frac{4}
{D-2}\Phi} g_{MN}
+ \epsilon_{\alpha\beta} \partial_\alpha X^M \partial _\beta X^N
B_{MN} + \Phi (X) R^{(2)}  ]
\label{first}
\eea
in which the latter appears as a renormalization counterterm.
 
\pr
The outline of this paper is as follows. In section 2 we
review the appearance of quantum fluctuations in
couplings of the world-sheet
field theory, as a consequence of higher-genus wormhole-like
configurations of the world sheet, emphasizing the r\^ole of
logarithmic operators~\cite{gur,kogmav}. We start section 3 by
       recalling the consistency of non-critical Liouville
       string theory with canonical quantization
of the space of string theories~\cite{emninfl}, and then introduce a
    prescription for the
path integral over string configurations.
A key element in our approach   is the treatment of the
Liouville field as a local renormalization scale~\cite{emn}.
The renormalization programme is developed in section 4, and
we demonstrate explicitly how the action (\ref{first})
emerges from
renormalization counterterms. We also derive the appropriate
second-order equation of motion for the $D$ brane.
Section 5 develops an
important application of this formalism to the problem
of black holes and information loss in string theory,
inspired by previous work in a two-dimensional pilot
example. We demonstrate~\cite{horava} the appearance of open strings
attached to the event horizon viewed as a $D$ brane~\cite{callan},
whose ends  carry Chan-Paton-like quantum numbers
related to the $W_\infty$ charges of the two-dimensional
string black hole. In our interpretation, it is the
``leakage" of these charges that leads to information
loss at the horizon~\cite{emn}, as perceived by an external observer.
Conclusions are presented in section 6.

\section{Formulation of string theory in resummed high-genus
world sheets}
\pr
In this section we introduce the formulation of the sum
over world-sheet configurations of a string theory.
We recall that, in the normal first-quantized treatment of
a string model at fixed genus, the background fields are
determined classically, and may
be regarded as coupling constants. However, the sum
over world-sheet topologies entails quantization of these
couplings, as we shall now argue.
Features of this quantization can be seen in the
non-perturbative matrix model formulation of
two-dimensional strings, in which the sum over
world-sheet topologies is handled via a discrete
triangulation of the world sheet. Similar conclusions
have been reached in the `wormhole calculus' approach
to the sum over topologically non-trivial space times
in four-dimensional quantum gravity~\cite{wormholes}.
We shall bring out the
analogies during the course of our development.
 
\pr
We start our development by reviewing relevant results
from $c=1$ Liouville theory~\cite{DDK}, which can be viewed as the
asymptotic spatial and temporal limit of the two-dimensional
black hole~\cite{witt}. We consider the extra logarithmic divergences
that arise, as we now discuss, from degenerate handles in the
summation over higher-genus Riemann surfaces. These may
in general be formulated as long, thin tubes connecting two
Riemann surfaces $\Sigma_{1,2}$. At the one-loop (torus)
order, the relevant configuration is a long, thin handle
attached to a sphere $\Sigma$. This may be regarded as a
world-sheet wormhole, in close analogy with conventional
wormholes in four-dimensional quantum gravity~\cite{wormholes}.
 
\pr
String propagation on such a world sheet may be described
formally by adding bilocal world-sheet operators ${\cal B}_{ii}$
on the world-sheet sphere~\cite{polchinski}:
\be
 {\cal B}_{ii} =\int d^2z V_i (z) \int d^2w V_i (w)
 \frac{1}{L_0+{\overline L}_0 -2}
\label{bilocal}
\ee
where the last factor represents the string propagator, $\Delta _S$,
on the handle, with the symbols $L_0, {\overline L}_0$
denoting Virasoro generators as usual. Inserting a
complete set ${\cal E}_{\alpha}$ of intermediate string
states, we can rewrite (\ref{bilocal}) as an integral
over the parameter $q \equiv e^{2 \pi i \tau}$, where
$\tau$ is the complex modular parameter characterizing
the world-sheet tube.
The string propagator over the world-sheet tube then reads
\be
\Delta _S \,=\,   \sum _\alpha   \int dq d{\overline q}
  \frac{1}{q^{1-h_\alpha} {\overline q}^{1-{\overline h}_\alpha}}
\{{\cal E}_\alpha (z_1)
\otimes (ghosts) \otimes
{\cal E}_\alpha (z_2) \}_{\Sigma_1 \oplus \Sigma_2}
\label{props}
\ee
where $h_\alpha, {\overline h}_\alpha $ are the
conformal dimensions of the states ${\cal E}_\alpha$.
The sum in (\ref{props}) is over all states
propagating along the long, thin tube connecting
$\Sigma_1$ and $\Sigma_2$, which are both equal to the
sphere in the degenerating torus handle case of interest.
As indicated in (\ref{props}), the sum over states must
include the ghosts, whose central charge cancels that of
the world-sheet matter theory in any critical string model.
 
\pr
States with $h_\alpha = {\overline h}_\alpha = 0$ may
cause extra logarithmic divergences in (\ref{props})
which are not included in the familiar $\beta$-function
analysis on $\Sigma$~\cite{paban}.
This is because such states make contributions to
the integral of the form
$\int dq d{\overline q} / q {\overline q}$
in the limit $q \rightarrow 0$, which represents a
long, thin tube. We assume that such states are discrete in
the space of states, i.e., they are separated from other
states by a gap. In this case, there are factorizable
logarithmic divergences in (\ref{props}) which depend
on the background surfaces $\Sigma_{1,2}$, e.g., the
sphere in the case of the degenerating torus.
 
\pr
The bilocal term (\ref{bilocal}) can be cast in the form of
a local contribution to the world-sheet action, if one
employs the trick, familiar from the wormhole calculus, of
rewriting it as a Gaussian integral~\cite{wormholes,paban}:
\be
  e^{{\cal B}_{ii}} \propto \int d\alpha
  e^{-\alpha _i^2 + \alpha _i\int V_i }
\label{wlocal}
\ee
where the $\alpha ^i$ are to be viewed as
quantum coupling constants/fields of the world-sheet
$\sigma$-model. Once it becomes apparent (\ref{wlocal})
that the couplings $\alpha_i$ must be treated as
quantum variables, it is natural to replace the factor
$e^{-(\alpha _i)^2} $ by a more general Gaussian distribution
of width $\Gamma$~\cite{wormholes,schmid2}:
\be
{\cal P} =
 \frac{1}{\sqrt{2\pi}\Gamma} e^{-\frac{1}{2\Gamma ^2}(\alpha _i)^2}
\label{gauss}
\ee
The extra logarithmic divergences associated with degenerate
handles that we mentioned above: $\propto \hbox{ln} \epsilon$
where $q \sim \epsilon \sim 0$, have the effect of causing the
width parameter $\Gamma$ also to depend logarithmically on
the cutoff scale $\epsilon$. Upon identification of the
cutoff scale with the Liouville field, and identifying the
latter with the target time variable, we infer that the
distributions of the couplings $\alpha_i$ become
time-dependent~\cite{aspects,schmid2}.
 
\pr
We have assumed in the above treatment that the
Virasoro operator $L_0$ may be diagonalized simply
in the basis of string states, with eigenvalues their
conformal dimensions $h_{\alpha}$:
\be
L_0 |{\cal E}_{\alpha}> = h_{\alpha} |{\cal E}_{\alpha}>
\label{diagonal}
\ee
However, this simple diagonalization may fail in the
presence of a non-trivial solitonic background~\cite{kogmav},
as
has been shown in a range of conformal models,
including flat-space
$c = 1$ Liouville theory~\cite{bilal}, two-dimensional
black holes~\cite{emn,kogmav}, and certain models of disorder in
condensed-matter physics~\cite{tsvelik}. In these cases, there
are logarithmic operators~\cite{gur}
associated with degeneracies
in the spectrum of conformal dimensions $h_{\alpha}$.
In particular, other operators may have dimension zero,
and hence be degenerate with the identity operator.
Examples in the flat-space
$c = 1$ Liouville theory are the two
conjugate dressings of the identity operator:
$e^{\sqrt{2}\phi}$, $\phi e^{\sqrt{2}\phi }$,
where $\phi$ is the Liouville field~\cite{bilal}.
 
\pr
In the presence of logarithmic operators, the OPE
features anomalous logarithmic factors, and there
is a non-trivial Jordan cell in which the Virasoro
operator is not diagonal. These logarithmic operators
therefore mix in string propagators, and may yield
additional logarithmic factors in the presence of a
degenerate handle, if the logarithmic operators carry
non-trivial weight in the sum over intermediate
states~\cite{kogmav}.
This is not the case in the flat-space
$c = 1$ Liouville string example
mentioned above, because the dressings of the identity
operator carry zero weight in the sum over intermediate
states. In such a case, the bilocal operator (\ref{bilocal})
yields distributions for the quantum couplings $\alpha_i$
which have constant widths $\Gamma$, as in standard
wormhole examples in four-dimensional quantum gravity with
fluctuating space-time topology. However, the logarithmic
operators in generic soliton backgrounds {\it do} have
non-trivial weights~\cite{kogmav},
and hence contribute to the scale-
and hence time-dependence of the distributions of the
couplings which we mentioned earlier as a possibility,
via divergences of double-logarithmic type:
$\int dq d{\bar q} \frac{1}{|q|^2}ln |q| \dots $,
in the region $q \sim 0$.
The double logarithm arises from the form
of the string propagator over the world-sheet tube
used in  (\ref{bilocal}), which,  in the presence of
generic logarithmic operators $C$ and $D$,
takes the form
\bea
\int dq d{\overline q}q^{h_C -1}  {\overline q}^
{\bar{h}_C -1} <CD|\left(
\begin{array}{cc}
1 & \ln q \\
 0 & 1
\end{array}\right)|CD><{\overline C}{\overline D}|\left(
\begin{array}{cc}
1 & \ln \bar{q} \\
 0 & 1
\end{array}\right)|{\overline C}{\overline D}>
\label{CDprop}
\eea
As was shown in ref. \cite{kogmav}, this mixing
between $C$ and $D$ states along degenerate handles
leads formally to divergent
string propagators in the amplitudes,
whose integrals have leading divergences of the form
\bea
\int \frac{ dq d{\overline q}}{ q {\overline q}} \left[ \ln q
\int d^2 z_1 D(z_1) \int d^2 z_2 C(z_2) +c.c.\right] \nonumber \\
 \sim (\ln^2 \epsilon)
\int d^2 z_1 D(z_1) \int d^2 z_2 C(z_2) + c.c.
\label{mixing}
\eea
giving a leading singularity $\propto \ln^2 \epsilon$. Besides this
term, there will be terms $\propto \ln \epsilon$, corresponding
to $\int d^2 z_1 D(z_1) \int d^2 z_2 D(z_2)$ and
$\int d^2 z_1 C(z_1) \int d^2 z_2 C(z_2)$ terms.
 
\pr
To see the origin of such non-trivial logarithmic
operators in a generic string-soliton case, consider
the scattering of a light string mode off a soliton,
including recoil effects.
The centre of mass of the soliton
must be treated as a collective coordinate~\cite{paban},
whose integration ensures momentum conservation: the centre
of mass
is shifted during a scattering process, as a
result of the recoil. This effect can be described
in the world-sheet $\sigma$-model path-integral
formalism by making a constant shift in the {\it spatial}
coordinates $X$ of space-time:
\be
         X^{\mu}(z, \bar{z})
 \rightarrow X^{\mu}(z, \bar{z}) + q^{\mu}; \qquad
 q^{\mu}=const
\label{three}
\ee
As discussed in refs. \cite{paban,kogmav},
such a deformation can be expressed as a total world-sheet
derivative
\be
       {\cal O}_I ={\cal N}
       (ghosts) \otimes \partial _\alpha \{ g^j(X)
       \frac{\delta}{\delta( \partial _\alpha X^I)}
V_j (\partial _\beta X ; {\cal J}) \} \equiv
{\cal N} (ghosts) \otimes \partial _\alpha J^\alpha _I
\label{five}
\ee
where ${\cal N}$ is a normalization factor, ${\cal J}$
are Kac-Moody currents,
and
$J^\alpha_I $ is a two-dimensional Noether current.
We consider as an example here the deformations
associated with translations (\ref{three}), which are labelled
by a latin index $I$ running over the (target-space)
soliton-background zero modes:
\be
   J_{\alpha, I} \equiv \frac{\delta S}{\delta(\partial _\alpha X^I)}
\label{current}
\ee
where $S$ is a world sheet action.
The ghost insertions in (\ref{five}), which are
of the form ${\overline c}c$,
must be included in order to treat correctly the
ghost zero modes, whose presence shifts the anomalous
dimension of the operator from (1,1) to (0,0).
\pr
It was shown in ref. \cite{kogmav} that
the (normalized) Zamolodchikov metric~\cite{zam}
corresponding to ${\cal O}$,
which is defined by~\cite{zam}:
\bea
&~&G_{{\cal O}{\cal O}} =4\pi \delta ^{IJ}
=|z|^4 <{\cal O}^I(z,{\bar z}),
{\cal O}^J (0) >= \nn \\
&~&{\cal N}^2  ~
|z|^4 < ({\overline \partial} J^I (z,{\bar z}) + c.c.)
(\partial {\overline J}^J(0,0) + c.c.) >
\label{six}
\eea
is {\it well-defined} and {\it finite}
if one postulates that in the soliton  background
of ref. \cite{paban} the currents
$J^I$ are not the usual $(1,0)$ currents,
but the {\it logarithmic} currents
obeying the following OPE~\cite{kogmav,gur,tsvelik}
\be
 <J^I(z)J^J(0)> = - \kappa G^{IJ} \frac{\ln |z|^2}{z^2}
\label{correctnorm}
\ee
with $\kappa$ a normalization constant.
We have used a non-trivial target-space background
metric $G_{IJ}$ above, so as to incorporate
general covariance, in a straightforward generalization of
the flat space result of ref. \cite{kogmav}.
It is now also straightforward
to verify that (\ref{correctnorm})
leads to a finite factor ${\cal N}^2$ and a well-defined
Zamolodchikov metric (\ref{six})
\bea
-\kappa {\cal N}^2
 |z|^4 \left(\partial^2 \frac{\ln |z|^2}{{\bar z}^2}
+ c.c.\right) =  2\kappa {\cal N}^2  = 4\pi; \nonumber \\
{\cal N}^2 = \frac{2\pi}{\kappa}
\label{correctN}
\eea
We contrast this with the conventional
case, assumed in ref. \cite{paban}, in which the currents
$J^I$ satisfy a standard non-logarithmic OPE
\be
 <J^I(z)J^J(0)> \sim \frac{\kappa G ^{IJ}}{z^2}
\label{seven}
\ee
The resulting `normalized' metric (\ref{six})
would have a logarithmically-divergent coefficient,
${\cal N}=log\epsilon$, and
the un-normalized Zamolodchikov metric tensor
would vanish logarithmically~\cite{paban}.
The importance of the finite metric (\ref{six})
for the central result of this work, namely
the derivation of the action (\ref{first}),
will become apparent in section 4.
 
\pr
We draw the reader's attention to the analogy between
the above discussion and the treatment of conifold
singularities in Calabi-Yau compactifications \cite{conifold}.
There, the metric in moduli space has an apparent
logarithmic singularity if massless quantum black holes are
not taken correctly into account. Here also, the inclusion
of new massless states associated with logarithmic operators
restores the regularity of the metric in theory space.
 
\pr
In the particular example developed above, the
extra logarithmic divergence (\ref{correctnorm}),
and hence the corresponding logarithmic operator
and extra divergence in the distribution function
(\ref{gauss}), result from the correct imposition
of momentum conservation via soliton recoil as in
(\ref{three}). This
is just one example of a transition between two
different conformal field theories, characterized
in this case by different locations of the string
soliton. Analogous logarithmic operators also appear
in the quantum treatment of the two-dimensional string
black hole, e.g., in the calculation of
instanton effects~\cite{yung,emn} - which reflect transitions between
black holes described by conformal field theories with
different central charges
and hence represent black-hole
decays, and in the
treatment of world-sheet monopole-antimonopole configurations -
which reflect the creation and annihilation of such a string
black hole~\cite{emn}. These are also examples of transitions between
different conformal field theories.
 
\pr
Analogous transitions, logarithmic operators and the
associated zero modes are clearly generic,
and their treatment requires going
beyond the conventional first-quantized approach to
string theory. As we have argued elsewhere \cite{emninfl},
and as we review at the start of the next section, they
necessitate a quantum treatment of the background
couplings. The generic soliton situation outlined
above corresponds to a saddle point in the eventual
string field theory path integral, and later in the
next section we make a proposal towards the formulation
of such a path integral based on a second-quantized
version of Liouville string theory. This
accommodates string solitons in a natural way, as well
as the eventual emergence of $D$ branes.
 
\section{Path-Integral Formulation of
Quantum String Theory Space }
\pr
We have argued above that the summation over world-sheet
topologies necessarily induces the
second quantization of string theory~\cite{emn}.
The correct formulation of this problem in generic
higher-dimensional string theory is far from complete.
In two-dimensional string theory, the summation over genera can be
performed exactly in the so-called matrix model
approach~\cite{matrix}, which
amouts to a discretization of the world sheet. There is
no rigorous proof that the continuum limit of such a model yields
the appropriate sum over genera of a world-sheet
continuum $\sigma$-model in an arbitrary
background. However, the summation over genera can be performed
in a flat space-time background, by using the
inverted harmonic oscillator potential of the $c=1$ matrix model.
From the point of view of Liouville strings, the interesting
case is that of fluctuating black-hole space-time
backgrounds. The incorporation of such backgrounds in the
matrix-model picture is still not very well understood,
though there are
attempts \cite{bhmatrix} to incorporate {\it fixed }
black-hole backgrounds
in the matrix-model framework, by utilizing
deformations of the inverted harmonic oscillator potential.
However, it is not yet known how to incorporate
a fluctuating (or, {\it a forteriori}, an evaporating)
black hole at the desired level of rigour.
Nevertheless, for the purposes of our discussion below,
we shall assume that the summation over genera in black-hole
string theory can be defined satisfactorily,
and we shall present a qualitative discussion
of its properties, based on a geometric description of theory space.
 
\pr
As we have argued elsewhere~\cite{emn}, the natural framework in
which to discuss the space of string theories is that of
renormalizable two-dimensional $\sigma$ models on the
world sheet, which is endowed with a
natural metric structure~\cite{zam}.
Motion through this theory space is determined at the
classical level by the renormalization $\beta$ functions
of the $\sigma$-model couplings, $\{ g^i \}$,
which are given by the
gradients of a suitable form of the Zamolodchikov $C$
function~\cite{zam}:
\be
\beta ^i = G^{ij} \partial _j C[g]
\qquad : \qquad G_{ij} = 2|z|^4 <V_i(z,{\bar z}), V_j (0)>
\label{gradient}
\ee
where the $V_i$ are the appropriate vertex operators
on the world sheet, describing the emission of
target-space states by the fields/couplings $g^i$.
This classical motion along Renormalization Group
trajectories may equivalently be expressed in terms
of an action principle, with the integrated Zamolodchikov $C$ function
serving as the action~\cite{emn,kogan}
$\Gamma [g] = \int dt L$:
\be
\Gamma[g]\,=\, \int dt C(g)\,
 =\, \int dt (p_i {\dot g^i} - H ) \nn
\label{action1}
\ee
The effective target-space Hamiltonian $H$ is defined by a
Legendre transformation of (\ref{action1}), and
provides the following representation of the
classical equation of motion:
\be
\partial _t \rho = -\{ \rho, H \} - G_{ij}\beta ^j
\partial _{p_i} \rho
\label{action2}
\ee
where the $\{\, , \, \}$ are classical
Poisson brackets,
$\rho (g^j, p_k)$ denotes
the matter density matrix,
the coordinates $g^i$ in string theory space are
identified with the $\sigma$-model couplings, and the vertices
$V_i$ are identified with the canonical conjugate
momenta $p_i$, upon identification of the Liouville scale
field with time~\cite{emn,kogan}.
\pr
We have argued in the previous Section that the summation
over higher genera necessitates quantization of the
background fields/couplings $g_i$. Canonical quantization
with the commutation relations:
\be
[g^i, g^j]=[p_i, p_j]=0 \qquad [g^i, p_j]= i\hbar \delta^i_j
\label{commut}
\ee
is possible only if the dynamical system obeys the
Helmholtz conditions~\cite{hojman}. If this is the case,
total time derivatives of functionals of $g^i$ are
then defined by commutators with the Hamiltonian operator $H$
\be
 \frac{d f(g)}{d t} = -i[ f(g), H ]
\label{qderiv}
\ee
and equation (\ref{action2}) becomes~\cite{emn,kogan}:
\be
\partial _t \rho = i[ \rho, H ] + iG_{ij}\beta ^j [g^i, \rho]
\label{action3}
\ee
where here and henceforth we set
$\hbar = 1 $.
\pr
We have shown elsewhere~\cite{emninfl} that non-critical strings
satisfy the Helmholtz conditions~\cite{hojman}, and here
we only recall some key features of the proof.
Formally, given a Newtonian-type equation of motion
for a generic mechanical system with coordinates $g^i$,
a sufficient condition for canonical quantization is the existence
of a non-singular symmetric matrix $w_{ij}$ and scale factor $Q$
with the property that \cite{hojman}
\be
w_{ij} ({\ddot g} ^j - Q {\dot g}^j ) =
\frac{d}{d\phi }(\frac{\partial L}{\partial {\dot g} ^i}   )-
\frac{\partial L}{\partial g ^i}
\label{conf17}
\ee
where the dots denote derivatives with respect to the
Liouville mode $\phi$,
for some function $L$ to be identified as the off-shell
Lagrange function of the system. If the appropriate
conditions for the existence of $w_{ij}$ and $Q$
are met, the solution for $w_{ij}$ is~\cite{hojman}
\be
  w_{ij} = \frac{\partial ^2 L } {\partial {\dot g ^i}
\partial {\dot g^j}}
\label{conf23}
\ee
 
\pr
We have shown that, in
the case of `off-shell' strings, discussed here,
the lagrangian $L (t)$
is just the Zamolodchikov $C$ function:
\be
 L(t ) = -\beta ^i G_{ij} \beta ^j
\label{conf25}
\ee
and $w_{ij}$ may be identified (up to a sign) with the
Zamolodchikov metric:
\be
 w_{ij} = - G_{ij}
\label{conf26}
\ee
In this case, the Helmholtz conditions are satisfied as a
result of non-trivial features of the Zamolodchikov
framework, in particular the facts that
$G_{ij}$ is only a functional of the $g ^i$
and not the $\beta ^i$, and that
it obeys the rescaling property
\be
     \frac{d}{d t} G_{ij} = Q G_{ij}
\label{conf27}
\ee
under the action of the on-shell Liouville time derivative $d/dt$
for some suitable scale factor $Q$,
which coincides in this case with the ordinary renormalization
group operator for the couplings $g^i$.
Due to the Liouville-renormalization-group invariance of
$Q$, (\ref{conf27}) implies a linearly-expanding \cite{aben}
scale factor for the metric in string theory space,
\be
    G_{ij} [t, g (t) ]= e^{Qt} {\hat G}_{ij} [t ,g (t) ]
\label{conf28}
\ee
where the notation ${\hat A}$ is used for
any Liouville-renormalization-group invariant
function $A$. This is exactly the form
of the Zamolodchikov metric in Liouville strings, as
discussed in ref. \cite{emninfl}.
\pr
The above results suggest that a
path-integral
formalism for the string theory space parametrized by the $g^i$
should be available, and we now
make a proposal for its construction.
The starting point for our proposal is the formalism
of Osborn and Shore \cite{osborn,shore}, in which the
background fields $g^i$ are allowed to depend explicitly on the
world-sheet coordinates, which we have applied
previously to the case of a fixed-genus Riemann surface \cite{emn}.
Here we generalize the approach of \cite{osborn,shore} to the
summation over world-sheet topologies:
motivated by the matrix model results, and replacing the
background fields $g^i$ by operators ${\hat g}^i$
in target space, we show later how the
form of the Zamolodchikov metric can be induced dynamically
when we perform a path integral over the variables $g^i$.
This approach assumes a Lagrangian formalism in coupling-constant
space, obtained by
integration over the conjugate momenta $p_i$ in the
target-space string effective action $\Gamma [g]$ (\ref{action1}).
Our proposal is to
write the path integral over string theories in the form
\be
    \int{\cal D}g^i e^{-\Gamma [g, A]} =
    \int {\cal D}g^i \int DX D\phi
 \delta (A - \int _{\Sigma} e^\phi ) e^{-(S^* +
 \int _{\Sigma} \{ g_i[V_i(X)] +\partial _\alpha g^i {\cal G}_{ij}
\partial _\alpha g^j + \Phi (g) R^{2} + \dots \})}
\label{pi}
\ee
where $S^*$ is a fixed point world-sheet action,
the $X$ are the target spatial coordinates,
and $\phi$ is the Liouville field, interpreted as the target time
coordinate. We have factored out the integral over
$A$, the global area of the world sheet, which serves
as a renormalization-scale evolution parameter, as described above.
The index $i$ includes a
summation over the target-space zero modes~\cite{emn},
which play a crucial role in the quantization of the $g^i$,
as we discussed
after (\ref{five}) in the previous section.
These should be not be confused
with the dependences on the space-time coordinates
$X$ and $\phi $, which we treat separately.
The world-sheet integral is over
the lowest-genus Riemann surface $\Sigma$,
since the summation over higher genera
is represented by the quantization of the couplings $g^i$,
treated here as path-integration variables.
Classical string vacua, which are described
by conformal backgrounds corresponding to fixed
points of the renormalization group flow, appear as
saddle points in the path integral
(\ref{pi}). This is a $\sigma$-model vision
of what a string field theory might be, which is not
necessarily complete, but captures the
features of interest to us for the derivation of (\ref{first}).
 
\pr
The interpretation of (\ref{pi}) is of a `string' moving in
an abstract space of possible theories, whose coordinates
are the background fields $g^i$. This is to be distinguished
on the one hand
from the usual motion of a string
in target space-time ($X$, $\phi$), and, on the other hand,
from the usual interpretation of the renormalization group
as the motion of a point-like particle in the space of
possible field-theoretical couplings.
Formally, (\ref{pi}) resembles a $\sigma$ model in theory
space, and we now demonstrate that the quantization induced
by the summation over higher genera endows this space with
its metric ${\cal G}_{ij}$, which appears in the integral
\be
    {\cal Z} = \int Dg^i e^{-\int _{\Sigma} \{ \partial _\alpha g^i
{\cal G} _{ij} \partial _\alpha g^i + \Phi (g) R^{(2)} +
C[g] \} }
\label{sigma}
\ee
The unconventional ${\cal G}_{ij}$-dependent terms
in (\ref{sigma}) appear as renormalization counterterms.
They would be
absent on a flat world sheet, since they vanish
as $\partial _\alpha g^i \rightarrow 0$, for $\alpha = z,{\bar z}$.
Their presence is essential, though, for the
consistency of the quantization in curved space~\cite{osborn}.
 
\pr
To see how the terms
in (\ref{sigma}) arise, we first recall~\cite{shore,osborn}
that
renormalizability implies the following
local renormalization group equation:
\be
{\cal D} {\cal L} \equiv
(\epsilon - {\hat \beta }  . \partial _g
- {\hat \beta } _\lambda . \partial _\lambda) {\cal L}
= 0  \qquad \lambda \equiv (\Phi, \Lambda )
\label{rge}
\ee
where ${\cal L}$ is a first-quantized world-sheet
$\sigma$-model lagrangian, and
the dot  includes integration over world-sheet
coordinates. The $\beta$ functions refer to the
structures appearing in (\ref{sigma}), with $\Phi$
denoting dilaton counterterms, and
$\Lambda$ representing
the extra counterterm involving ${\cal G}_{ij}$.
These arise from the explicit world-sheet coordinate
dependence of the renormalized couplings $g^i$
on the curved world sheet :
\bea
{\hat \beta }_g^i &=&\epsilon g^i + \beta ^i (g)  \qquad ; \qquad
{\hat \beta }_\Phi = \epsilon \Phi + \beta ^\Phi   \qquad ; \qquad
{\hat \beta }_\Lambda = \epsilon \Lambda + \beta ^\Lambda
\qquad ; \nn \\
\beta _\Lambda &=& \frac{1}{2}
\partial _\alpha g^i \chi_{ij} \partial ^\alpha g^j
\label{betas}
\eea
where $\chi _{ij} $ in (\ref{betas}) is defined via
\be
(\epsilon- {\hat \beta}^k \partial _k
) {\cal G}_{ij} - (\partial _i {\hat \beta }^k
) {\cal G}_{k j} - (\partial _j {\hat \beta}^k)
{\cal G}_{ik} = \chi _{ij}
\label{chi}
\ee
It can be shown~\cite{osborn} that $\chi _{ij}$
is related to the coincidence limit of the Zamolodchikov
metric~\cite{zam}, defined through the two-point functions
of vertex operators $V_i$.
 
\pr
In order to discuss the antisymmetric-tensor term in (\ref{first}),
we also need to address the appearance of
torsion terms in coupling constant space,
Such terms can be non-zero on the boundaries
between `patches' in theory space.
The formal inclusion of such terms
presents no essential
difficulties~\cite{osborn}. It should be noted that
in the off-shell corollary of the Zamolodchikov $C$-theorem~\cite{zam},
which relates off-shell variations of the string effective action
in target-space, $\Gamma $,  to
the $\sigma$-model $\beta$-functions,
such coupling-constant-space torsion terms appear
explicitly~\cite{osborn}:
\be
    \frac{\delta }{\delta g^i} \Gamma [g]
      = \chi _{ij} \beta ^j
+ (\partial _i {\cal W}_j - \partial _j {\cal W}_i)\beta^j
\label{torsion}
\ee
where the ${\cal W}_i$ are
well-defined functions of renormalized couplings,
related to total derivatives on the world sheet
in the expression for the trace $\Theta $
of the $\sigma$-model stress tensor
in terms of the
(renormalized) couplings/fields $g^i$~\cite{osborn}:
\be
  \Theta = \beta ^i [V_i] + \partial_\alpha Z^{\alpha}
\qquad ; \qquad Z^\alpha = {\cal W}_i \partial ^\alpha g^i
\label{ttrace}
\ee
The bracket in (\ref{ttrace})
denotes a normal product with respect to the renormalization
group flow variable that we identify with time.
Due to their total-derivative form, the counterterms ${\cal W}_i$
play a non-trivial r\^ole in the case of world sheets
with boundaries, as appear when we incorporate
open strings in the spectrum of physical theories.
These and related issues will be discussed
in the next section.
\pr
Having identified the most relevant structures in the path
integral (\ref{pi}), we now discuss consistency conditions
that it must satisfy. Integrating (\ref{pi}) over $X$ and the
Liouville conformal factor $\phi$, we arrived at
the path integral (\ref{sigma}). This
makes sense as a quantum field theory
on the world-sheet when one imposes
the absence of the conformal anomalies, which
correspond to the vanishing of the $\beta$ functions associated
with the quantities ${\cal G}_{ij}$, $\Phi (g)$ and $C[g]$,
considered as couplings of a $\sigma$ model formulated
over a target manifold parametrized by coordinates  $g_i$.
Thus, their $\beta$ functions should not be confused with the
ordinary renormalization-group coefficients pertaining
to the target-space fields $g^i$.
Moreover, conformal invariance
of any  $\sigma$ model on a target manifold implies the
vanishing, not of the usual renormalization-group
$\beta$ functions, but of the Weyl
anomaly coefficients, which are invariant under
target-space diffeomorphisms \cite{shore}.
In our case, it is known that a renormalization-scheme
change corresponds
to diffeomorphisms in the coordinates $g^i$, in the sense of
local field redefinitions~\cite{review}.
 
\pr
The conformal invariance of the last term in (\ref{sigma}) is
ensured by taking into account our use of the global world-sheet
area $A$ as an ordinary renormalization-group
scale, which implies that it
coincides with the usual renormalization-group invariance
of the Zamolodchikov function $C[g] \sim \Gamma [g]$.
Conformal invariance of the middle `dilaton' term in
(\ref{sigma}) restricts the form of $\Phi$ in a manner which
does not concern us here. Finally, the conformal-invariance
conditions for the first term in the action in (\ref{sigma}),
namely the Zamolodchikov
metric background, read
\be
   \frac{d}{dt } {\cal G} _{ij} = \nabla _{(i} W_{j)}
\label{weylmetr}
\ee
We now recall
(\ref{chi}), which we write in the form
\be
\chi _{ij}
= {\cal G}_{ij}^{(1)} + {\cal L}_{\beta } {\cal G}_{ij}
=\frac{1}{2}
 ( \partial _t + {\hat \beta} ^k \partial _k ){\cal G }_{ij}
 +  \nabla _{(i} (\hat \beta ^k {\cal G}_{j)k})
 - {\hat \beta }^k \Gamma _{kij}
\equiv {\cal D} {\cal G}_{ij} + \nabla _{(i} V_{j)}
\label{crist}
\ee
where $(ij)$ denotes symmetrization of indices,
${\cal L}_{\xi}$ is the Lie derivative
with respect to a coordinate transformation $\xi $, and
$\Gamma _{kjl}$ is the Christoffel symbol for the
`metric' ${\cal G}_{ij}$ . Thus,
up to diffeomorphism and connection terms, which
can be removed by appropriate scheme choice, the right-hand side
of (\ref{crist}) corresponds to the usual renormalization group operator
acting on ${\cal G}_{ij}$. The scheme for which
the connection $\Gamma _{ijk}$ vanishes corresponds to
a `normal coordinate choice', which will be adopted in
the definition of our path integral (\ref{pi})\footnote{Connection
terms are related to
contact terms of moduli operators, and the reader might worry
that such schemes cannot be easily adopted in string theory,
where moduli fields play an important r\^ole. However,
in our case there will be no such problem.
Our normal coordinate choice
implies vanishing connection with respect to the
metric ${\cal G}$, which is different from the
Zamolodchikov metric $G_{ij}$ that is directly related
to the moduli fields.}.
 
\pr
We now explore the consequences of (\ref{weylmetr}), and
check its consistency with standard $\sigma$-model
lore. To this end, we first observe
from the middle expression in (\ref{crist}) that
a change in renormalization scheme relates $\chi _{ij}$
to the residue of the simple pole in $\epsilon$
of ${\cal G}_{ij}$. From arguments
in the string literature, we also know~\cite{mavc}
that there exists a scheme which relates
$\chi_{ij}$ to the {\it coincidence limit} of the original
Zamolodchikov tensor~\cite{zam}.
From our conformal invariance condition
(\ref{weylmetr}), then, it follows that
\be
    \chi _{ij} = \nabla _{(i}(W_{j)} - V_{j)})
\label{cons}
\ee
As we shall argue now, it will be necessary that
\be
 W_i - V_i =\nabla _i C [g]
\label{gauges}
\ee
where $C[g]$ is identified with the Zamolodchikov function,
up to an irrelevant constant.
To see this, we use the definitions
\be
[O_i (x)] = \frac{\delta S}{\delta g^i (x)}
\label{vertex}
\ee
for the renormalized vertex operators $[ O_i ]$, where $S$ is the
$\sigma$-model action on a world-sheet with coordinates $x$. From
this, we can easily derive the following expression :
\be
<[O_i](x)[O_j](0) > = \frac{\delta ^2 \Gamma [g] }{\delta g^i(x)
\delta g^j (0)} - <\frac{\delta }{\delta g^i (x)}[O_i ] (0) >
\label{doubleder}
\ee
The first term takes a K\" ahler form, whilst the second is
non-zero only in the coincidence limit $x \rightarrow 0 $,
and in fact is proportional to a world-sheet $\delta$ function,
since $\delta g^i (x)/\delta g^j (y) = \delta ^{i}_j
\delta ^{(2)} (x-y)$.
Such local
terms do not contribute to the Zamolodchikov metric~\cite{mavc},
either because one can always defines the latter at large scales
$|x| \rightarrow \infty$, or else because such local terms are
subtracted to ensure the renormalization-group invariance
of the pertinent two-point function. Details
have been given in ref.~\cite{mavc}.
As we have argued there, one can always find a class of renormalization
schemes for which the generating functional of connected string
amplitudes $\Gamma [g]$ can be transformed to the
Zamolodchikov
function
$C[g]$. This can be achieved by
redefinitions that involve only the metric $G_{ij}$
and not the connection in coupling-constant space.
Thus, such schemes are always operational in the
`normal coordinate' system~\footnote{The subtleties
with the moduli fields in certain target-space supersymmetric
string theories
must be dealt with in  the way
described above.} in string theory space. Therefore,
(\ref{cons}) is consistent with (\ref{doubleder})
if and only if the choice (\ref{gauges}) is satisfied.
 
\pr
This completes our discussion of the conformal invariance
of (\ref{sigma}). In turn,
this justifies the path-integral ansatz (\ref{pi}):
the latter can indeed be regarded as a generalized
$\sigma$ model with quantized couplings, as
suggested previously \cite{emninfl}.
 
\pr
As a corollary to the above proof, we remark that
if we complexify the coupling constant
space, as is appropriate for the $N=2$ superconformal
theories of interest to string theory
with non-trivial duality symmetries~\cite{ooguri},
then (\ref{gauges}) implies a K\" ahler form
for the metric (\ref{cons}). As we have seen above,
this K\" ahler form of the Zamolodchikov metric
appears
consistent with its standard definition
in terms of the divergences of two-point functions
which are not removed by conventional renormalization.
This will be important for section 5, when we shall present
arguments in favour of a fundamental r\^ole played
by two-dimensional target-space strings in the
path integral (\ref{pi}).
 
\pr
We close this section with a few important comments.
We stress once again
that the tensor ${\cal G}_{ij}$ {\it is not}, in general,
the same as the Zamolodchikov metric:
$G_{ij} \propto \chi _{ij}$. Rather,
the
renormalization-group derivative of ${\cal G}_{ij}$
coincides with $\chi_{ij}$ in certain schemes.
In our quantum version of the
theory space, it is $ {\cal G}_{ij} $
that appears as the `metric' of the $\sigma$-model
on the manifold $g^i(x)$. Thus, the effect of the
quantum corrections due to higher genera is that
the `string' should be regarded as a particle described by
a $\sigma$ model with target space metric
${\cal G}_{ij}$ and {\it not } $G_{ij}$.
One can argue, though, that there exists
a class of renormalization
schemes - compatible with {\it canonical} quantization
of the coupling/fields $g^i$ - in which
the Zamolodchikov metric has the structure
\be
\chi _{ij} =  G_{ij} =
Q^2{\cal G}_{ij}  + \dots
\label{qprop}
\ee
where $Q^2=\frac{1}{3} (C[g]-25)$ , with $C[g]$ the Zamolodchikov
$C$-function, which in the renormalization
scheme chosen above
is equivalent to the
effective action of the low-energy
field theory. The $\dots$ denote terms that
either refer to moduli fields (exactly marginal
deformations)
of the string,
or to antisymmetric tensor backgrounds, which
drop out of the symmetrical
counterterm involving ${\cal G}_{ij}$ in (\ref{pi})
(c.f. (\ref{cterm}) below).
The moduli-dependent terms are important
to ensure passage from one background to another
in a smooth way, but their details
can be ignored for our
present purposes.
The relation (\ref{qprop})
follows from a combination of
the off-shell corollary (\ref{torsion})
of the
$C$-theorem~\cite{osborn}, which allows an identification
of $\chi _{ij}$ with $G_{ij}$, the relation
(\ref{crist}),
and the conditions
(\ref{conf27},\ref{conf28})
for
canonical quantization.
We argue in the next section that one finds in this class of
schemes consistent string soliton solutions,
which need sources for their support~\cite{gibbons}.
The latter solutions play an important r\^ole in understanding
the structure of string theory, especially from the point of view
of recent developments in the context of duality symmetries~\cite{conifold}.
\pr
 
\section{$D$ branes and solitons in string field theory}
\pr
We now show how the above path-integral framework (\ref{pi})
leads to an action of the form (\ref{first}), in which
a {\it target}-space string effective lagrangian is combined with
a {\it first}-quantized {\it world-sheet} action, the latter
describing the coupling of the string background to a test
string source~\cite{gibbons}. We start by considering the
local counterterm
\be
\Delta S = \int _{\Sigma}
d^2\sigma
 \partial _\alpha g^i {\cal G}_{ij} \partial ^\alpha g^j
\label{cterm}
\ee
in the exponent of (\ref{pi}). As discussed in the previous section,
this arises from the
consistent renormalization of the $\sigma$-model action
on a curved world sheet, in the context of the summation
over higher genera. This approach~\cite{osborn} relied on
couplings $g^i$ becoming {\it local} functions of the world-sheet
variables $z,{\bar z}$. This dependence was allowed to be
arbitrary in ref. \cite{osborn}, whereas it was taken to be
through the Liouville mode~\cite{emn,shore} in the
first-quantized approach to the Liouville string.
However, when one proceeds to second quantization of the
Liouville string, the $g^i$ also depend on $z,{\bar z}$
via the target-space $\sigma$-model
fields $X^M(z,{\bar z})$.  This dependence
is in addition to any target-space zero-mode
dependence included among the indices $i$ of
the $\sigma$-model couplings  $g^i$~\footnote{In
fact, the dimensionality
of the $X^M(z,{\bar z})$ may be different from that of
the original target space we started with, and in this
way one can dynamically generate target-space dimensions
starting
from low-dimensional target-space strings, e.g. the two-dimensional
black hole~\cite{witt}. More discussion on this
point will be presented in the next section, in connection
with a conjectural fundamental r\^ole of two-dimensional
target-space strings in the path integral (\ref{pi}).}.
 
\pr
We use this implicit dependence on the world-sheet coordinates
via the target-space coordinates $X^M(z, {\bar z})$ to rewrite
the counterterm (\ref{cterm}) in the path integral (\ref{pi}):
\be
\Delta S\,=\,
\int d^2\sigma \partial _\alpha X^M
\partial ^\alpha X^N
\frac{\delta g^i}{\delta X^M}
{\cal G}_{ij}
\frac{\delta g^i}{\delta X^N}
\label{source1}
\ee
We use now the
fact that the metric ${\cal G}_{ij}$
is related through (\ref{qprop}) to $\chi _{ij}$, which
is connected in a given scheme~\cite{osborn,mavc}
to the coincidence limit of the Zamolodchikov metric~\cite{zam}
$G_{ij}$,
which is in turn related to two-point
functions of vertex operators in the generalized world-sheet
$\sigma$ model. Applying the $\sigma$-model equations of
motion, we can rewrite (\ref{source1}) in the form
\be
\Delta S
\,=\,\int d^2\sigma \partial _\alpha X^M
\partial ^\alpha X^N lim_{z \rightarrow w} |z-w|^4
\frac{1}{Q^2}<\partial^\alpha J_\alpha^M (z,{\bar z})
\partial^\beta J_\alpha^N (w,{\bar w})>
\label{source2}
\ee
The currents $J^M$ appearing in (\ref{source2}) are the same
Noether currents as in (\ref{current}), which have the
unconventional OPE (\ref{correctnorm}) in a soliton background,
which we relate in turn
to the finite Zamolodchikov metric (\ref{correctN}).
By virtue of (\ref{correctnorm}), then,
the expression (\ref{source2})
 may in turn be re-expressed in
terms of the target-space metric $G_{MN}$
\be
\Delta S\,\propto \, \int d^2\sigma \partial _\alpha X^M
\partial ^\alpha X^N G_{MN} (X)
\label{source}
\ee
The absence of explicit
logarithmic divergences in (\ref{source}) is consistent
with the conformal-invariance conditions (\ref{weylmetr})
of the string-theory-space path integral (\ref{pi}).
Notice that this construction works {\it only}
for string backgrounds with logarithmic currents.
For conventional currents $J^M$ satisfying
the non-logarithmic OPE (\ref{seven}), the
unrenormalized Zamolodchikov metric ${\cal G}_{ij}$ vanishes
logarithmically~\cite{paban}, and hence there is no associated
contribution to the path integral (\ref{pi}).
\pr
This discussion can be extended to antisymmetric-tensor backgrounds,
which are also of interest for string solitons and membranes~\cite{duf}.
In their presence, there are additional structures which lead to an
antisymmetric tensor coupling in (\ref{source}), and
the relevant renormalization-group
counterterms in (\ref{pi}) take the form
\be
    \int d^2\sigma \epsilon^{\alpha\beta}
\partial_\alpha g^i (
\partial _i {\cal W}_j -
\partial _j {\cal W}_i ) \partial_\beta g^j
\label{torsct}
\ee
The counterterm (\ref{torsct}) is dictated
by the $\sigma$-model renormalization group
relations (\ref{torsion}), (\ref{ttrace}), which
imply that the metric in theory space of the
stringy $\sigma$-model contains a torsion (antisymmetric)
part.
Replacing $\partial _\alpha g^i $
by $\partial _\alpha X^M \frac{\delta g^i}{\delta X^M}$,
with the $X^M$ depending on the world-sheet coordinates,
one obtains an antisymmetric-tensor source term at
the point
$X^M$ in the
effective target-space action
\bea
&~&
   \int d^2 \sigma \epsilon^{\alpha \beta}
   \partial _\alpha X^M \partial _\beta X^N
B_{MN} (X)  \nn \\
&~&
B_{MN} = \partial _N \Lambda _M (X)
- \partial _M \Lambda _N (X) \equiv
\frac{\delta {\cal W}_i }{\delta X^M}\frac{\delta g^i}{\delta X^N}
\label{antisym}
\eea
For regular string background fields $g^i$, such terms do not
lead to non-trivial source terms, as the corresponding antisymmetric
tensor fields can be gauged away. However, for
points in $g^i$ space that
have singularities at $X^M$ (e.g., solitonic string backgrounds
such as black holes), $\Lambda (X)$ may be singular
at $X$, thereby leading to a non-trivial field strength
for the antisymmetric field $B_{MN}$ defined as in (\ref{antisym}).
The presence of logarithmic operators
in such backgrounds
guarantees the smooth character of the antisymmetric-tensor
coupling at the source point
$X$, in a way similar to the graviton coupling (\ref{source}).
 
\pr
Finally, we also observe that terms in the path
integral (\ref{pi}) that couple to
the world-sheet curvature $R^{(2)}$
through `dilatons' $\Phi [g]$ in theory space $\{ g^i \}$
lead to a dilaton $\Phi (X)$ contribution to the target space
effective action, $C[g]$, and hence
to the string-source $\sigma$-model
action in order $\alpha '$.
\pr
Combining the above results, we find that,
in a low-energy effective-action treatment of $C[g]$,
the complete exponent of (\ref{pi}) in a soliton background
can be written as
\bea
&~&{\hat S} = C[g] +~{\rm string~source~terms} =  \nn \\
&~& \frac{1}{g_s^\chi} \int d^DX\sqrt{G}e^{-2\Phi} [R +
4(\nabla \Phi )^2 - \frac{1}{12}H_{MNR}^2
+ O(\alpha ')] + \nn \\
&~&\frac{1}{4\pi \alpha ' }
\int d^2\sigma [
\partial _\alpha X^M \partial ^\alpha X^N G_{MN} (X)   +
\epsilon^{\alpha\beta} \partial_\alpha X^M \partial_\beta X^N
B_{MN} + \Phi (X) R^{(2)} ]
\label{membrane}
\eea
where the metric $G_{MN}$ is the $\sigma$-model
metric, $g_s $ is the string coupling constant
\be
g_s  \equiv
 e^{<\Phi>}
\label{gees}
\ee
and  $\chi$ is the Euler characteristic
of the world-sheet manifold. For world-sheets with the topology
of the sphere, $\chi =2$, whilst for those with the
topology of a disk, as appropriate for a theory containing
open strings in its spectrum, $\chi =1$.
Note that the appearance of the string coupling constant
in front of the effective-action part of (\ref{membrane})
is a result of the non-trivial $Q^2$ normalization factor
appearing in (\ref{qprop}), which enters the relevant
expression for the counteterm (\ref{cterm}).
Eq. (\ref{membrane}) describes
the action for a macroscopic string (soliton) solution
coupled to a string source~\cite{gibbons}.
In order to pass to
a renormalization scheme where the
Einstein term in the target-space effective action
assumes the canonical form (\ref{first}), one has to rescale the
metric
$G_{MN} \rightarrow
e^{\frac{4}{D-2}\Phi} G_{MN} \equiv g_{MN}$.
In terms of $g_{MN}$, the action assumes the form (\ref{first}).
It is a highly non-trivial fact that solutions to the
above coupled system of world-sheet and target-space
objects are found. From our point of view such solutions
correspond to
stationary points of the path integral
in string `theory space' (\ref{pi}).
\pr
The
explicit appearance of
the string coupling constant
in (\ref{membrane})
motivates the definition of a new scale parameter
\be
\alpha_s' \equiv \alpha'/g_s^\chi
\label{newscale}
\ee
Using a system of units based on the new
fundamental scale $\alpha _s' $ (\ref{newscale}),
the natural cut-off scale
that appears in the low-energy target-space
effective action in (\ref{first})
becomes $\alpha _s' g_s^\chi $.
In the case of $D$ branes~\cite{dbrane}, which are associated
with open strings, $\chi =1$, and the new scale
is $\alpha _s' g_s$.
\pr
For a four-dimensional string, the scale $\alpha _s' g_s$
appearing in the rescaled target-space action
should be identified with the
Planck scale of quantum gravity.
Such a scale, involving the string coupling constant,
has been conjectured to exist in quantum solitonic
string theory and in $D$-brane theory~\cite{shenker},
and has also appeared in the two-dimensional stringy
black-hole example
of ref. \cite{emn}, where it has been related to the
entropy of a decaying stringy black hole~\cite{emn4d}.
We remark that the flat-space
scattering of soliton strings and $D$ branes, both
systems characterized by a no-net-force condition,
cannot probe scales shorter than $\alpha _s'$~\cite{bachas}.
Thus, it is still not known how one can probe scales of order
$\alpha _s' g_s$ using weakly-coupled strings with $g_s < 1$.
However, it may well be that such scales are probed by
superscattering $\nd{S}$-matrix
elements of soliton strings and $D$ branes
in highly-curved (black-hole) target space times.
In this respect we
note that the field redefinition of the metric field
which allows a passage from (\ref{membrane}) to (\ref{first})
has been made in an essentially off-shell effective action.
Usually, such local redefinitions leave the on-shell $S$-matrix
elements invariant, and therefore do not affect the low-energy
physics. However, in the presence of a string source or
macroscopic string, and in general in string
backgrounds such as black holes, etc., where an
$S$-matrix cannot always be defined, local
redefinitions may  well
affect the physics, as a result of the off-shell character
of the target space action in (\ref{membrane})~\cite{emn,tseytl2}.
 
\pr
At this point we address the key question how
open string states     appear in our formulation of quantized
string field theory (\ref{pi}), which would allow
    $D$-brane structures~\cite{dbrane} to emerge.
To see how the approach leading to (\ref{pi})
naturally incorporates world-sheet boundaries, which are
associated with open string states,
we consider
world-sheet dependences of the couplings $g^i$
such that $\partial _\alpha g^i =0$ in some regions
of the two-dimensional surface appearing in the integrand of
(\ref{pi}). In such cases,
the imposition of appropriate conditions
on the boundaries separating these regions from the
rest of the Riemann surface leads to
interesting saddle-point solutions of (\ref{pi}) such as
$D$ branes~\cite{dbrane}. The reason is that
in the presence of a hole in the
world sheet there is a boundary contribution coming from the
counterterm (\ref{cterm})
\be
   \int _{\partial \Sigma} <g^iV_i \partial ^\alpha J_{\alpha,M}>
\partial _n X^M
\label{gf1}
\ee
where $\partial _n$ denotes derivatives normal
to the world-sheet boundary $\partial \Sigma$.
On the other hand, from the torsion counterterms (\ref{torsct})
one gets
\be
   \int _{\partial \Sigma} <g^i \partial _{[i} W_{j]} \frac{\delta g^j}
{\delta X^M}>
\partial _\tau X^M
\label{gf2}
\ee
where $[\dots]$ denotes antisymmetrization of the respective
indices, and
$\partial _\tau$ denotes derivatives tangential
to the boundary $\partial \Sigma$.
By appropriate choice of the functions $X^M$
on the boundary, one may induce in this way
the $U(1)$ gauge fields that appear in $D$-brane
theories~\cite{dbrane}. For instance, in Type II
superstring theories, if one chooses Neumann boundary conditions
for some coordinates of the target manifold,
$X^i$, while the rest, $X^I$, obey fixed Dirichlet~\cite{dbrane}
boundary conditions,
the term (\ref{gf1}) leads to transverse excitations of the
$D$ brane, which are
represented by insertions of the operator
\be
\int _{\partial \Sigma} A_I \partial _n X^I
\label{gf3}
\ee
On the other hand, the term (\ref{gf2}) leads to
longitudinal excitations, corresponding to an internal
gauge field associated with the gauge invariance of the
antisymmetric tensor field: $B_{MN} \rightarrow
B_{MN} + \partial _{[M}\Lambda _{N]}$:
\be
\int _{\partial \Sigma} A_i \partial _\tau X^i
\label{gf4}
\ee
The gauge field operators (\ref{gf3},\ref{gf4}) induced
in this way may then be absorbed in redefinitions (scheme choices)
of target-space $U(1)$ gauge fields, which appear naturally
in any theory involving open strings\footnote{It
is to be understood that one uses in such cases the formal
technology of $\sigma$ models with boundaries~\cite{mcavity},
appropriately extended in the framework of (\ref{pi}).}.
 
\pr
The construction above makes evident the similarity of $D$ branes
to string solitons, and $D$ branes have indeed been
viewed in the literature~\cite{dbrane,dbrsc}
as sources for closed string fields obeying
the standard vanishing $\beta$-function conditions
of critical string theory outside the $D$ brane.
A fixed boundary
is analogous to a fixed string source, as studied
above, as far as the breaking of certain target-space symmetries
is concerned. By an appropriate choice of
boundary conditions~\cite{dbrane},
the whole boundary conformal field theory is mapped onto a single
point in target space $\{ X \}$.
However, the integration over boundaries (i.e. over $\{ X \}$)
in (\ref{pi}), dictated
by the renormalization-group approach~\cite{osborn},
should restore these symmetries.
Logarithmic operators
should also play a r\^ole in this case, by analogy with
the restoration of translational or diffeomorphism
invariance through soliton-recoil/back-reaction
effects in the case of macroscopic strings~\cite{kogmav}
(\ref{membrane}) or stringy black-holes~\cite{emn,mnbrain}.
Indeed, while this paper was
close to completion, an interesting work~\cite{periwal}
appeared in which logarithmic operators describing $D$-brane
recoil have been explicitly constructed in the spirit of
ref. \cite{kogmav}.  For instance, in the simplified
example of a $0$ brane (point particle)
discussed in ref. \cite{periwal},
the logarithmic operators
describing recoil
are given by  the boundary contribution
\be
 V_{recoil} = v_I \int d^2 \sigma  \partial _\alpha (X^0 \Theta (X^0)
\partial ^\alpha X^I)
\label{recoil}
\ee
where $v_I$ is the (center-of-mass)
velocity of the $0$ brane,
the position of its centre of mass as a
function of time $X^0$ being $v_I X^0$, and $\Theta (X^0)$
is the Heaviside step function.
The time $X^0$ obeys the standard
Neumann boundary conditions on the boundary
of the world sheet, whilst the $X^I$ obey fixed
Dirichlet boundary conditions,
corresponding to the location of the $0$ brane,
regarded as a collective coordinate.
According to the analysis of ref. \cite{periwal},
the operator (\ref{recoil}), viewed as a deformation
of the $0$ brane on the disc, cancels logarithmic
divergences of amplitudes for
matter excitations in the $0$-brane background on
the annulus, in complete analogy to the closed-string
soliton case, discussed in ref. \cite{kogmav} and
in section 2. This interpretation of (\ref{recoil})
as solitonic recoil passes the consistency check
of yielding a soliton mass $\propto 1/g_s$,
where $g_s$ is the string loop coupling constant.
Moreover, it is consistent with the
above unified approach to strings and $D$ branes from the
non-critical-string
path integral (\ref{pi}) point of view.
The deformation (\ref{recoil}) has the general
form of the contribution (\ref{gf3}) to the
$D$-brane boundary state~\cite{dbrsc}, and may be thought of
as corresponding to a Lorentz boost with time $X^0$.
In our approach in which the Liouville
field is interpreted as time~\cite{emn,aben,dvn},
the step function appearing in (\ref{recoil})
is a reflection of the monotonic flow that results
from the semigroup nature of the renormalization group~\cite{emn,kogan}.
\pr
Further formal evidence for the consistency of such an
identification
is provided by the
exact
flat-space
renormalization-group equation
for the wave functional of the $D$ brane, postulated~\cite{rey}
on the basis of Polchinski's
extension of the Wilsonian approach~\cite{liu}.
Let $\Psi [X^I]$ be the wave functional of the $D$ brane,
which is related by
$Z = \int dX^I \Psi [X^I]$
to the partition function $Z$
of the first-quantized version
of the $D$ brane, formulated on world sheets with boundaries and
resummed over genera.
The wave functional is, thus, given by~\cite{rey}
\be
 \Psi [X^I] \sim e^{S_D}~~;~~
   S_D \equiv \sum _{h=1}^{\infty} g_s^{h-2} S_{(h)}
\label{wfdb}
\ee
where $h$ denotes the number of holes (the sum over handles has been
suppressed
in the above notation), and $S_D$ sums up
all one-particle irreducible connected world-sheet amplitudes $S_{(h)}$
whose boundaries are mapped onto the world-volume of the
$D$ brane.
 
\pr
The formal
exact renormalization-group equation for a $D$ brane in flat space
reads~\cite{rey}
\be
\epsilon \partial _\epsilon \Psi [X^I]
=\frac{1}{2}\int _{\partial \Sigma} \int _{\partial \Sigma}
\epsilon \partial _\epsilon G_{IJ} [X^I, X^{J}]
\frac{\delta ^2}{\partial X^I \partial X^J }\Psi [X^I]
\label{polchinsk}
\ee
where $\epsilon$ is the world-sheet renormalization scale,
related, for example, to the degeneracy of handles as discussed
in section 2, and
$G_{IJ}[X^I,X^J]$ denotes the two-point function
of the $\sigma$-model fields $X^I$, obeying
Dirichlet boundary conditions. We expect, on the basis of
our analysis in section 2, that such two-point
functions contain $log \epsilon$ divergencies
due to the zero-mode contributions
\be
\hbox{Lim}_{\epsilon \rightarrow 0}
 G_{IJ}[X^I,X^J] \sim - |{\overline X}^I|^2
\delta ^{IJ}
 log \epsilon
\label{zero}
\ee
where the ${\overline X}^I $  denote $c$-number
zero modes of the transverse coordinates,
which are independent
of the world-sheet coordinates of the $D$ brane.
For instance, in the case of the $0$ brane
discussed in ref. \cite{periwal},
it is straightforward to verify
(\ref{zero}) by computing the two-point function
of the operator (\ref{recoil}) describing translation
of the transverse coordinates of the $0$ brane. Concentrating
on the $X^0$ part, sufficient for demonstrating (\ref{zero}),
 one obtains for the zero-mode contributions
\bea
&~& <v_I X^0 \Theta (X^0) v_J X^0  \Theta (X^0) >_{zero~mode} \sim  \nn \\
&~&
|v_I^2| \delta ^{IJ} <X^0 \Theta (X^0)(z=i\frac{1}{2}\epsilon)
>^2 =
|v_I {\overline X}^0 \Theta ({\overline X}^0)|^2 \delta _{IJ}
(\int \frac{dr}{r^2}
\epsilon^{-r^2})^2 \sim \nn \\
&~& - |v_I {\overline X}^0 \Theta ({\overline X}^0)|^2 \delta _{IJ}~log \epsilon
\label{log}
\eea
where we have used an integral representation
for the $\Theta$ function, and
taken into account the fact that,
near the boundary of the world sheet $z=i\frac{1}{2}\epsilon$,
the v.e.v of the recoil operator
(\ref{recoil}) is divergent~\cite{periwal} as a result of the zero modes,
and
needs to be regulated by a proper-time cut-off $\epsilon$.
 
\pr
Equipped with (\ref{zero}), equation (\ref{polchinsk}) can
be verified by
a direct computation of the infinities
of the annulus amplitudes due to
the massless zero modes (logarithmic operators)~\cite{kogmav}
associated with the translation of the
transverse coordinates of the $D$ brane and related
to recoil~\cite{periwal,kogmav}.
Such divergences are target-space infrared in nature,
as we mentioned in section 2 and in the above discussion,
and are associated with
bilocal operator insertions on the disk amplitudes
of the translation operator  (\ref{gf3}):
\be
      V_{T} \equiv \int _{\partial \Sigma} A_I (X^0, ... X^p) \partial _n X^I
= A_I \frac{\partial}{\partial X^I}
\label{transl}
\ee
for a $D$ brane
with $10 - p$ collective coordinates,
i.e., with
$p$ coordinates $X^0, ..., X^p$  obeying Neuman boundary conditions.
To leading order in the string coupling constant $g_s$,
only the disk and annulus amplitudes contribute.
From (\ref{wfdb}), (\ref{transl}) it is, then,
immediate to write down the
cut-off dependence
of the $D$-brane wave function boosted with velocity $V$~\cite{rey}
\be
     \Psi_V [X^I] = e^{-log\epsilon \frac{1}{2} g_s \nabla ^2 _{X^I}}~\Psi [X^I]
\label{eq}
\ee
which is consistent with the identification of $-log \epsilon$
with a Euclideanized target time  $T$.
Equation (\ref{eq})
is an approximate expression which yields
a diffusion-like equation of motion
for the wave functional of the $D$ brane~\cite{rey}
\be
\frac{d}{d T} \Psi_V = -\frac{1}{2M_D} \nabla _X^2 \Psi _V
\label{schr}
\ee
where
$M_D \propto 1/g_s $ is the $D$-brane mass.
\pr
 
\pr
However, the above analysis is restricted to a $D$ brane
moving in flat target space, and is modified in an
essential way when curved backgrounds are taken into account,
as we now discuss. We emphasize first that,
in the presence of curved target-space backgrounds,
not all boundary conditions are compatible
with conformal invariance. For instance,
in the presence of a linear-dilaton background~\cite{aben}
fixed Dirichlet boundary conditions are not conformal
invariant~\cite{dbrane}. To restore conformal invariance,
one needs to
incorporate certain extra boundary operators~\cite{Mli}.
From our non-critical string approach, this phenomenon
is compatible with the quantum string nature of the
$D$ branes. The modification of the boundary conditions
to restore conformal symmetry
is the analogue of Liouville dressing in our non-critical
string approach~\cite{emn}.
A possible connection between such boundary modifications and
logarithmic operators in the boundary, that induce
interactions with the bulk,
might be anticipated in view of the above discussion.
This is in close analogy to the case of
chiral edge currents of the Hall systems in condensed matter
physics~\cite{hallc2}.
There, the presence of a boundary
condition on the edge current induces
extra local interactions of $\delta$-function type
in the potential of the
quantum-mechanical Scr\"odinger equation. Such interactions
exhibit non-trivial renormalization-group
scaling properties~\cite{hallc2}.
A rigorous formulation of such issues in the context of (\ref{pi})
is under study at present.
\pr
Nevertheless, from the discussion above, we can infer
some useful properties of the time evolution
of $D$ branes with fixed Dirichlet boundary conditions
in such non-conformal curved backgrounds. We take the
point of view that, instead of modifying the boundary conditions so
as to ensure conformal invariance~\cite{Mli}, one should
rather view this type of problem with fixed non-conformal
boundary conditions on the collective coordinates of the string soliton
as an effective non-critical string problem requiring
dressing by the Liouville field, interpreted
as a local renormalization-group scale to be identified with
target time~\cite{emn,kogan}. In this way, we derive the
correct equation of motion of the $D$ brane wave function
in the spirit of our picture (\ref{pi}),
which incorporates $D$ branes as saddle-point solutions of
ordinary string theories in soliton backgrounds.
Indeed, a generic $D$-brane world-sheet action can always be viewed as
an open-string action in appropriate gauge-field backgrounds,
with all the coordinates of the string obeying standard
Neumann boundary conditions~\cite{dbrane,bachas}.
In this interpetation, the two terms in (\ref{gf3},\ref{gf4})
can be interpreted as stanard open-string gauge background fields,
upon the replacement of the normal derivative $\partial _n X^I$ by
a tangential derivative $\partial _\tau X^I$ ,
with the $X^I$ obeying standard Neumann boundary conditions.
The formal reason for this equivalence is that the following
relation is valid on the world-sheet boundary:
\be
<\partial _\sigma X^I (\tau) \partial _\sigma X^J (\tau ') >_{Dirichlet} =
- <\partial _\tau  X^I (\tau) \partial _\tau X^J (\tau ') >_{Neumann} = \frac{2\alpha ' \delta ^{IJ}}
{(\tau - \tau ')^2}
\label{cond}
\ee
Using the abovementioned equivalence,
the target-space effective action of the $D$ brane on a flat target space
is then given by the Born-Infeld Lagrangian~\cite{yost}
\be
{\cal L}_{BI} \propto \sqrt{-det(\eta_{\mu\nu} +
2\pi \alpha ' F_{\mu\nu})}, \qquad
; \qquad \mu,\nu = 1, \dots 10
\label{born}
\ee
where $\eta_{\mu\nu}$ is a Minkowski target-space flat metric,
and $F_{\mu\nu}$ is
the field strength of the background
(Abelian) gauge field appearing in (\ref{gf3},\ref{gf4}).
Going back to the Dirichlet picture,
the above string effective action yields
the standard Nambu-Goto
world-volume action of a $p$ brane~\cite{dbrane,bachas}, expressed
in terms of Dirichlet coordinates:
\be
  {\cal L}_{p-brane} \propto \sqrt{-det(\eta_{\alpha\beta} +
\partial _\alpha X^M \partial_\beta X_M)}
\qquad ; \qquad \alpha,\beta = 1, \dots p \qquad
M=p+1, \dots , 10
\label{pbrane}
\ee
where  $\eta_{\alpha\beta}$ is the flat world-volume metric.
 
\pr
In the case of curved backgrounds for the world volume,
conformal invariance
may be lost, as discussed above~\cite{Mli,dbrane}
for the fixed Dirichlet case,
in the sense that non-conformal deformations of the
$\sigma$ model appear. In the string picture, the
effective Born-Infeld Lagrangian depends also on the gravitational
target-space metric field, and in general on non-conformal
background fields $g^i$. As in our two-dimensional
black-hole picture~\cite{emn}, criticality is restored by
Liouville dressing. The deviation from criticality
in a curved background is parametrized by
by $Q^2 \propto C[g] -25$, where the Zamolodchikov function
exceeds $C[g] > 25$, so that the theory is critical and the
Liouville field $\phi$ has negative time-like metric.
As discussed elsewhere~\cite{emn,tsc},
this approach leads to a second-order
renormalization-group equation for the
Liouville-dressed couplings $\lambda ^i (\phi)$~\cite{emn}:
\be
{\ddot \lambda }^i(\phi)   + Q {\dot \lambda }^i (\phi) =-\beta ^i(\lambda)
\label{seor}
\ee
where the dots denote differentiation with respect to $\phi$,
and the $\beta ^i$ are the flat-world-sheet $\beta$-functions.
Notice that the friction term proportional to $Q$ in (\ref{seor})
can be abosrbed into a `mass-shift'
for the mode $\lambda ^i$~\cite{aben}
but, contrary to the case of ref. \cite{aben}, the mass is real here,
despite the Minkowskian signature of the
field $\phi$ when interpreted as target time~\cite{emn}:
\be
  [\partial ^2 _\phi  - \frac{1}{4}Q^2 ]e^{-\frac{1}{2}Q\phi} \lambda ^i = \beta ^i(e^{-\frac{1}{2}Q\phi}\lambda ^i)
\label{target}
\ee
Coming now to our main ojective, namely the
derivation of an equation of motion for the
$D$-brane wave functional $\Psi [X^I]$,
in our intepretation of time as a Liouville renormalization-group scale,
we observe that, because
of the abovementioned equivalence (\ref{born},\ref{pbrane})
of the $D$ brane
to an open string in background fields, one can represent
$\Psi $ as
\be
    \Psi [X^I] = e^{-C[\lambda]}
\label{psieff}
\ee
where
the Zamolodchikov function
$C[\lambda]$ is
the generating functional of connected Green functions,
which depends on the various non-conformal backgrounds.
In the dual $D$-brane picture it corresponds to $S_D$
in (\ref{wfdb}).
The source of the violation of conformal invariance is
the soliton background, which is characterized by
the Dirichlet boundary
conditions of its collective coordinates or, equivalently
in a closed string framework, by the presence
of a conformally-non-invariant
macroscopic string source (\ref{first}), as in the
discussion of this paper.
Let us then denote by $Q^2 $ the effective vacuum energy of the soliton,
which will include its mass $M_D$~\footnote{In the presence of other
background deformations, there will be other background-dependent
terms.}. It is known~\cite{tsc}
in the non-critical string picture that
the effective central charge obeys a second-order equation
near a fixed point:
\be
 {\ddot C}[g, \phi]
 + Q [g,\phi] {\dot C} [g,\phi] = -
 \beta^i \frac{\partial}{\partial \lambda ^i} C[\lambda] + O[{\dot \lambda}^i \frac{\partial^2 C[\lambda]}{\partial
\lambda ^i \partial \lambda ^j} {\dot \lambda }^j]
 \le 0 ~{\rm for}~C \ge 25
\label{ctheorem}
\ee
where
we have made use of the
conventional
flat-world-sheet
$C$-theorem~\cite{zam,emn,tsc}:
$
 -\beta^i \frac{\partial}{\partial \lambda ^i} C[\lambda] \le 0 $.
From (\ref{psieff}), (\ref{ctheorem}),
it is then straightforward to generalize the renormalization-group
equation (\ref{polchinsk}) to the Liouville case, and
derive an evolution equation
for the wave functional of the $D$ brane:
\be
[\partial ^2 _{X^0} -\frac{1}{4} Q^4 - \nabla^2_x -
\dots ]e^{-\frac{1}{2}Q^2 X^0}\Psi =0
\qquad Q^2 \ni M_D   \propto \frac{1}{g_s}
\label{liouv2}
\ee
where $X^0 \equiv \phi /Q$ is a covariant renormalization-group scale, which
behaves as a Minkowskian evolution parameter, and may thus be identified
with the target time~\cite{emn,kogan},
and
the dots denote possible additional
terms including interactions among the $D$ branes,
which are not relevant for our purposes here.
The above `derivation'  of (\ref{liouv2})
exploits the equivalence of $D$ branes
to some closed solitonic string backgrounds, which follows from our
path-integral ansatz (\ref{pi}).
In  this respect, equation (\ref{liouv2})
is a direct consequence of the fact that the $D$ branes
are nothing other than saddle-point solutions of (\ref{pi}).
In standard $D$-brane/string language,
this equivalence arises via certain
duality symmetries~\cite{dbrane}.
It would be interesting, as a further consistency
check of our approach, to derive (\ref{liouv2}) in an independent way,
extending the derivation of (\ref{schr})~\cite{rey} directly
to the case of curved world-brane volumes.
 
\pr
We close this section with some comments on previous
attempts to derive the action (\ref{membrane}) from
standard $\sigma$-model theory.
It has been argued, correctly, in ref. \cite{tsemem}
that the action (\ref{membrane}) can be considered the result
of world-sheet resummation in $\sigma$-models.
The non-conformal invariance of the $\sigma$-model involved
is traced back to the loop corrections to the $\beta$ functions.
In particular a heuristic derivation of this action
has been given by attaching a world-sheet wormhole
operator (\ref{bilocal}) to a $\sigma$ model on a
Rieman surface of lower genus. This is similar in spirit
to our analysis above. The crucial difference in our
approach, however, is the proper incorporation of
recoil effects on the soliton background
via logarithmic operators on the world sheet~\cite{emn,mnbrain,kogmav}.
In our opinion, this provides a
more complete treatment of the problem.  As
we have seen above, in the version (\ref{pi})
of string-theory-space quantization which
we advocate in this work, derived
from a computation of such recoil/back-reaction effects
in a $\sigma$-model language,
the crucial source/background
coupling terms for string solitons
originate from a particular renormalization-group
counterterm (\ref{cterm}). This term is also
responsible for a dynamical determination of
the metric in theory
space via the conditions (\ref{weylmetr}).
Such terms have not been considered in the previous
literature on the subject~\cite{tsemem}.
Although our derivation above was formal,
it has the merit of providing an explicit demonstration
that membrane theories are intimately connected with a
quantum target-space string theory. Moreover,
our quantization scheme (\ref{pi}), which is
consistent with the
non-critical nature of the (quantum) strings involved
and our identification of time in string theory
with the Liouville mode~\cite{emn},
allows for a proper transition between the
quantum and classical worlds in target space-time,
via decoherence effects
associated with the solitonic backgrounds~\cite{aspects}.
 
\section{Application to Black Holes and Information Loss}
 
\pr
Basing our arguments on the two-dimensional black-hole
prototype, we suggested some time ago that the Hawking-Bekenstein
entropy of a black hole could be understood as the
number of string states~\cite{emn4d}, in agreement
with the previous suggestion of ref. \cite{kaln} based on
string duality. This suggestion has been repeated
recently in the context of $D$-brane studies, where the
black-hole entropy may be interpreted as the number of
open string states that terminate on the horizon~\cite{callan}.
We have gone further, and argued that, to the extent
that these black-hole string states are unobserved, even
unobservable, they necessarily lead to microscopic
information loss~\cite{emn}. The amount of this information loss
was related, in our interpretation, to the loss of
internal $W_\infty$ quantum numbers carried by the two-dimensional
stringy black hole~\cite{emn,emnsel1,emnsel2}.
In this section, we use the development
of the previous sections to
discuss further aspects of the $D$-brane approach to
black hole information problem, indicating how
features of our previous two-dimensional
analysis may be carried over into the $D$-brane approach.
 
\pr
We start by recalling that open strings also appear
naturally in the two-dimensional black-hole model,
when one takes correctly into account its
asymptotic space-time geometry~\cite{witt}.
The reason is that, like all maximally-extended
Schwarzschild geometries in General Relativity,
the original two-dimensional black-hole model
contains two asymptotically-flat domains.
These may be reduced to a single asymptotic
domain by applying the familiar
orbifold technique of modding out by a $Z_2$
factor~\cite{horava}. This construction leads to
open unoriented string states attached to the
orbifold singularity at the origin. In this two-dimensional
model, such states
are necessarily {\it discrete} delocalized states,
and therefore cannot be detected in local scattering
experiments. This exemplifies the loss of information
to which we have drawn attention previously~\cite{emn}.
An interesting duality exists~\cite{horava} that maps the
model to a theory of open strings in a black hole
background.
 
\pr
The above discussion refers to the space-time singularity
at the core of the black hole.
As has been discussed elsewhere, this singularity may be
represented as a topologically non-trivial
world-sheet monopole configuration~\cite{hallc,emn}.
In this picture, it has been argued~\cite{hallc}
that the horizon of the black hole may also be
represented as a world-sheet (anti)monopole.
The picture outlined in the previous paragraph
therefore implies
that the horizon of the two-dimensional black hole should
also admit open-string states~\footnote{The explicit duality
symmetry~\cite{giveon} that maps the
singularity of the two-dimensional stringy
black hole~\cite{witt} onto the horizon offers
further support for this picture.}, in accord with the
intuitive picture of ref. \cite{sussk}, according to which
a closed string state falling into the horizon of the black hole
may lie partly inside and partly outside
the horizon.
 
\pr
There are similarities, but also important differences,
between our previous two-dimensional (2D) picture
and this emerging higher-dimensional (HD) one,
which deserve attention. In the 2D picture,
the open-string states are discrete~\cite{horava} and
non-propagating, and carry the gauged $W_\infty$ `quantum hair'
of the black hole~\cite{emn}. In the HD picture,
one starts from a string living in some higher-dimensional
space time with some dimensions compactified.
This leads to modes~\cite{callan} winding
around a compact dimension,
characterized by discrete momenta.
In the dual description which is
appropriate for a $D$-brane representation
of the horizon of the black hole~\cite{callan},
such modes are associated with open-string states whose ends are
confined to the $D$ brane, and also have discretized momenta.
Such states may be either left- or right-movers.
These two-dimensional
substructures play an important r\^ole~\cite{callan}
in the excitation of the various
black-hole states that correspond to a given set of
macroscopic classical hair (mass, charge and angular momentum)~\cite{wilc}.
Such excitations are obtained by considering
the excitation of oppositely-moving pairs
of such open strings, with their end points
attached to the black-hole horizon~\cite{callan}.
This corresponds to the intuitive `splitting'
picture developed in ref.~\cite{sussk} for
a closed string state falling inside the
horizon of a string black hole. The
counting of such states matches in a certain
perturbative limit~\cite{callan} the classical
Hawking-Bekenstein law.
The existence of discrete-momentum, essentially two-dimensional,
structures on the horizon of the black hole, which are responsible
for the degeneracy of the black-hole states,
is very similar to the situation in the 2D picture
of~\cite{witt,emn,horava}.
 
\pr
However, an important feature of the 2D picture,
which in our opinion does not yet seem to have been
take properly into account in the emerging HD picture, is the
peculiar nature of the quantum hair carried by the
{\it discrete} open-string states of the two-dimensional
black hole. This is essential to any
discussion of the maintenance or otherwise of
quantum coherence in black-hole decay~\cite{emn,emn4d}.
Estimates of entropy production
along the lines in ref.~\cite{callan},
although very important, do not address the nature of
the precise mechanism of the information transfer
from propagating to non-propagating (open-string)
states of the ($D$-brane) black hole.
On the other hand, two-dimensional strings are known
to possess $W_\infty$ symmetries with
a coherence-preserving property that maintains
unitarity in the temporal evolution {\it at the
fundamental string level}~\cite{emn}.
This admits a more-or-less conventional conformal-invariant
world sheet description~\cite{emn,chaudh}, provided that one takes
into account the discrete modes of the two-dimensional string.
 
\pr
However, the treatment of quantum-gravitational fluctuations
necessitates further analysis when one considers
the time evolution of a propagating closed-string mode
in the presence of a stringy black hole~\cite{emn}.
Since a low-energy observer is restricted to making
local measurements, which are insensitive to the
non-local discrete modes which carry information via their
$W_{\infty}$ quantum numbers, the time evolution in target space
does not take the familiar unitary form. The analogies between
the 2D and HD pictures outlined above, in particular
the fact that two-dimensional stringy black holes~\cite{witt}
also admit~\cite{horava} open string structures after orbifolding,
suggests that analogous modifications of unitary evolution
may also occur in the HD case. This motivates the study of
open string states - especially their $W_{\infty}$ quantum numbers -
that we carry out below.
 
\pr
Before embarking on this analysis, we first discuss how the
non-critical two-dimensional black-hole string model may be
embedded in a higher-dimensional theory within
our approach. One may introduce observable matter fields
as perturbations on the black-hole $\sigma$ model. One
may also consider tensoring the black hole with an
initially-decoupled flat-space-time non-critical
string of the type introduced in \cite{aben}:
\be
S^{nc} = \int d^2z [\delta_{MN} \partial X^M {\overline \partial}X^N
+ Q X^M R^{(2)} ]
\label{abenstring}
\ee
where $Q^2=\frac{1}{3} (D-25)$ in the case of a bosonic string, which we
consider for simplicity. Once the
integration over the backgrounds $g^i$ in (\ref{pi})
is considered,
the two theories couple to each other, and one encounters the
typical Liouville-dressing problem of a non-critical string
with a {\it single} Liouville coordinate. The latter represents
Minkowski target time if the central charge of the theory
is higher than 25~\cite{aben}, as is the case
of the two-dimensional black hole string perturbed
by world-sheet instanton deformations~\cite{emn}.
These may be used to represent the integration over
the global string modes of the string black hole
which is necessarily made by any low-energy observer.
The r\^ole played by two-dimensional strings is
fundamental in this approach, since they characterize the
global, non-propagating, string modes which are also
present in higher-dimensional theories. Their presence creates
an essentially `unobservable' environment
in quantum gravity theories~\cite{emn}, leading to the
loss of microscopic loss of quantum coherence
engendered by the loss of the $W_{\infty}$ charge information
which we now discuss.
 
\pr
It is essential for this purpose to understand
in more detail the nature of the hair
carried by a black hole, extending our previous analysis
of the $W$ hair that appears in the
2D case to the emerging HD treatment based on $D$ branes.
This extension must take account of the special nature of
the $W$ hair, which is associated with a gauged phase-space
$W_\infty$ symmetry~\cite{wgauge,emn}. The corresponding
$W$ currents are non-local~\cite{bakas}, and hence
the $W$ charges resemble `quantum' hair \`a la Aharonov-Bohm,
rather than conventional classical gauge hair. Indeed, we
have shown~\cite{emn,emnsel1,emnsel2}
how the $W$ hair is carried by discrete
non-propagating modes in the 2D model, and how it may
in principle be measured by (an infinite number of)
Aharonov-Bohm phase measurements.
\pr
To see how this picture may be reflected in the HD
approach, we consider the physics of the black-hole
horizon, viewed as a membrane with open strings attached
to it. The effective action appropriate for an observer at
infinity is given by (\ref{first}), as derived from the string
path integral (\ref{pi}) in Section 3, for the background-field
configurations appropriate for the description of a stringy
black hole.
In our approach, the
string world sheet has two space-like dimensions, with target time
appearing as the Liouville mode. There is therefore
a one-to-one map between the lowest-genus world-sheet
configuration and the black-hole horizon.
The world-sheet of a string source
can loop the horizon of the black hole~\cite{wrapkog},
and this is a quantum effect from an effective target-space point
of view.
An open string
appears as a handle attached to the horizon at defects in the
world sheet, each of which we may surround by a small circle,
whose radius we call $L$. The emergence of such
defects from the path integral (\ref{pi})
appears consistent with the above picture.
The physics of this boundary may
be described by a $c=1$
matrix model on a circle of radius $L$.
This is motivated by the fact that
in our approach we view the world-sheet of the string
as representing a (smooth Euclidean) target space
of another string theory~\cite{mgreen}. This
allows the representation
of the world-sheet as a $c=1$ matrix model, which is a theory
of free fermions on a Euclidean two-dimensional space time~\cite{matrix}.
One can introduce
a `time' for such configurations by identifying it with the
Liouville mode of the full black-hole string theory that has
Minkowskian signature.
In this sense, the free fermions that live on the
closed lines representing the defects
constitute a `string theory' living on a
(1 + 1)-dimensional target space-time.
Then, the
whole picture, consisting of a world sheet and a Liouville
mode,
represents a membrane whose boundaries are $2d$ Liouville theories.
Such constructions are known explicitly to exist~\cite{ashworth},
and are based on gravitational Chern-Simons
theories on three-dimensional manifolds with boundaries.
In this work we base our discussion on their mathematical
consistency.
\pr
It has recently been argued \cite{poly} that such a $c=1$
matrix model is equivalent non-perturbatively to a
two-dimensional pure Yang-Mills theory of $U(N \rightarrow \infty)$
(or, under suitable restrictions, $SU(N \rightarrow \infty)$)
type on a torus, with the large-$N$ limit appearing as
$L \rightarrow 0$. The world-sheet boundary circles are
closed circles in target space, with target time the second
dimension. If $g$ is the gauge coupling constant, the
corresponding $c=1$ string coupling constant is \cite{poly}
$4/g^2 L N$. We are indeed interested in the limit where the radius
$L$ of the defect in the world sheet vanishes, which we combine
with the large-$N$ limit so that the product $NL$ remains finite.
The above discussion then tells us that a weakly-coupled string
theory formulated on the boundary of the world-sheet defect
is equivalent to strongly-coupled gauge field theory
formulated at the ends of the open string.
\pr
The `tachyon' of the Das-Jevicki Hamiltonian~\cite{das}, which appears
in the continuum limit of the $c=1$ matrix model as a
momentum mode of the $c=1$ string living on the degenerating
$L \rightarrow 0$ world-sheet boundary, appears in the
$SU(N \rightarrow \infty)$ gauge field theory as a discrete
winding mode. This model may be described in terms of free
fermions
$\psi$ located at positions $\theta _i
: i = 1,2,...N \rightarrow \infty$ living
on the $L \rightarrow 0$ circle, with Hamiltonian
\be
     H= -(\frac{g^2 L}{2}) \sum _{i=1}^{N} \frac{\partial ^2}
{\partial \theta _i^2} \qquad ; \qquad 0 \le \theta _i < 2\pi
\label{fermiham}
\ee
Fermionization is a consequence of the appearance
of the Vandermonde determinant \cite{matrix} in the wave function:
$\Delta \equiv \Pi _{i < j} sin\frac{1}{2}(\theta_i - \theta _j )$.
One can, therefore, construct a second quantized effective field theory
of (\ref{fermiham}) which is that of the $c=1$
matrix model in the free-fermion $\psi $ representation~\cite{matrix}.
Bosonization yields the Das-Jevicki Hamiltonian~\cite{das}
on a circle.
\pr
In terms of the fermion fields $\psi$, the generators of the
$W_{1+\infty}$ symmetry are found by first decomposing
the fermion field on the circle into two oppositely-moving
Weyl fermions $\chi ^{(\pm)}$ which describe the relevant degrees
of freedom in the vicinity of the Fermi surface of the model,
as appropriate for an effective field-theory limit,
\be
\psi (\theta, t) = \frac{1}{\sqrt{L}} [
e^{ip_F (\theta - p_F t)}
\chi ^{(+)}(\theta - p_F t) +
e^{-ip_F (\theta + p_F t)}
\chi ^{(-)}(\theta + p_F t) ]~;
\qquad p_F=\frac{\pi (N -1)}{L}
\label{chiral}
\ee
with $p_F$ the Fermi momentum.
For {\it each} chirality, the relevant generators of the
associated
$W_{1+\infty}$ symmetry  are then given by~\cite{matrix,lerda}
(we omit the chirality index $\pm$ for simplicity)
\be
   V_n^m = \int _0^{2\pi} d\theta \chi ^\dagger (\theta)
:~e^{-in\theta} (i\partial _\theta )^m~: \chi (\theta)
\label{gener}
\ee
where $:~~:$ denotes normal ordering,
which satisfy
the commutation relations of a quantum $W_{1+\infty}$
algerba with central extensions~\cite{pope}
\be
  [V_n^i, V_m^j] = (jn-im)~V_{m+n}^{i+j-1}
+ q(i,j,m,n)~V_{m+n}^{i+j-3} + \dots +
\delta _{m+n,0} c(n,i,j)
\label{walg}
\ee
Above, the functions $q$ represent quantum corrections~\cite{pope},
and $c$ denotes the central extensions.
\pr
The $SU(N \rightarrow \infty)$ Yang-Mills gauge theory,
which the above free-fermion problem (\ref{fermiham})
is equivalent to~\cite{poly},
is the counterpart in configuration space of the
$W_\infty$ symmetry (\ref{walg})
of the two-dimensional strings,
which may be viewed as a gauge symmetry in
the two-dimensional
phase space~\cite{wgauge}.
The appearance of the $SU(N)$ symmetry allows us to
introduce more-or-less conventional
Chan-Paton factors at the ends of the open string,
where it couples to the horizon membrane.
This procedure may be viewed, in some sense, as
the $D$-brane extension of the fermionic representation
on the world-sheet boundary of the conventional Chan-Paton factors of
open strings~\cite{marcus}.
These $SU(N \rightarrow \infty)$
gauge charges are the counterpart of the
$W_\infty$ charges~\cite{floratos}
which we identified in the 2D model as playing
a key role in the black-hole information problem~\cite{emn}.
\pr
In the 2D case, we have argued that these $W_\infty$ charges
label the different black-hole states, and that the inability
of a low-energy observer performing local experiments to
measure these $W_\infty$ charges is responsible for the loss
of information in a black-hole environment. We have, moreover,
related~\cite{emn} the loss of information directly to the ``leakage"
of $W_\infty$ charge to unobservable modes in  our effective
non-critical string treatment of low-energy dynamics. The
above analysis demonstrates explicitly how this picture may be
translated into the emerging HD picture. It has been pointed
out~\cite{callan}
that the entropy of the black hole may be related to the
multiplicity of open-string states living on the horizon
surface, and we have demonstrated above
that these carry the HD analogues of the $W_\infty$ charges.
We believe that a full treatment of the low-energy dynamics
outside the black-hole horizon will confirm the loss of
information through ``leaks" of the black-hole quantum
numbers via these open string states.
\pr
We conclude this section with a brief comment on the
treatments of the black-hole singularity within the 2D and
HD approaches. It is thought that the black-hole
singularity has a topological nature, and there is indeed
an explicit
topological description of the singularity in the 2D case
in terms of a model with twisted $N=2$ supersymmetry~\cite{witt,eguchi}. This
model can again be mapped into a $c=1$ string theory on a circle,
which may also be described in terms of fermions, interacting
this time via potential terms. There is again a duality
symmetry that maps this model into a $U(N)$ gauge theory with
fermionic matter in some representation $R$ which is related
to the string-model interactions. Consider, for example, the
simplified case of a $U(N)$ gauge theory with a heavy colour
source, represented by a time-like Wilson line,
whose action is~\cite{poly2}:
\be
  \frac{1}{4} \int d^2 x F_{\mu\nu}^2
+ \int dt {\overline \Psi} (i\partial _t - g A_0(x=0)^a T_R^a
+ M) \Psi
\label{heavy}
\ee
where $M$ is the mass of the heavy colour source, and the $T_R^a$
are the generators of the gauge group $U(N)$ in the representation
$R$. This problem can be mapped into a $c=1$ string theory
which may be written in terms of interacting fermions
$\psi _i $, located at positions $\theta _i,
i =1, 2, \dots N$ on a circle,
with Calogero-Sutherland type interactions~\cite{poly2}:
\be
H=-\frac{g^2 L}{2} \sum _{i}^{N} \frac{\partial ^2}
{\partial \theta _i ^2} + \frac{g^2 L} {8}
\sum _{i \ne j} \frac{2C_{2m} /(N-1) + 2 L_i ^a L_j ^a }{sin^2
\frac{1}{2}(\theta _i - \theta _j)}
\label{calogero}
\ee
where the $L_i^a$ are generators of the $U(N)$ group, and
$C_{2m}$ is the quadratic Casimir for the $m$-fold symmetric
representation of $SU(N)$.
The limit $L \rightarrow 0$, appropriate for our
two-dimensional string description, can be taken
in (\ref{calogero}), at the expense of the appearance of
a strongly-coupled ($g^2 \rightarrow \infty$) gauge field
theory living on a space point, which is the limit of a
circle of vanishing length.
This simplified example manifests the complexity
arising when a specific type of heavy matter
is incorporated in the two-dimensional Yang-Mills theory.
The point we wish to emphasize here is that,
in the effective field theory description
of the Calogero-Sutherland model,
one can construct explicitly a $W_{1+\infty}$
spectrum-generating symmetry algebra, in an analogous way
to the free-fermion case~\cite{lerda}.
This suggests that such a symmetry also plays a key r\^ole
at the black-hole singularity. However,
the situation there is much more complicated,
and the pertinent analysis
still incomplete. Technical complications arise from
the twisted space-time supersymmetry~\cite{eguchi}
 which gives rise
to an enhanced topological super-$W_\infty$~\cite{emn} at the
the black-hole singularity, which possesses  a bosonic
$W_\infty \times W_\infty$ subsymmetry. We expect that the
emerging HD picture will also contain such topological
features and an enhanced symmetry at the singularity, but
the investigation of these points lies beyond the scope
of this paper.
 
\section{Summary and Conclusions}
 
\pr
In the approach followed in this paper, critical string theory emerges as
an ``effective" description, analogous to the chiral
lagrangian description of QCD in the conventional
low-temperature vacuum containing quark and gluon
condensates. The spectrum of this bosonized version of
QCD contains solitons, that can be identified as baryons \`a la
Skyrme. As one approaches from below the critical temperature
for the hadron/quark-gluon transition in QCD, the quark
condensate vanishes, as does the baryon mass. Above the
critical temperature, the appropriate description of QCD is
in terms of light quarks. As has been suggested previously,
black holes in string theory are analogous to
baryons~\cite{emn,emnsel1,emnsel2}, and
the vanishing of their masses is also believed to herald the
appearance of a new vacuum. A complete treatment of QCD must
go beyond the effective lagrangian treatment. Likewise,
a complete treatment of string
theory must include black holes and the non-perturbative
transitions they induce, and its natural description may well not
be in terms of ``effective" strings.
 
\pr
We have discussed in this paper an approach to
the formulation of string field theory which is based on
our non-critical Liouville string description of quantum
fluctuations in the space-time background. As we have
discussed in section 2, these arise inevitably when one
considers higher-genus configurations of the string world sheet.
Particularly important in this respect are configurations with
degenerate handles, which induce quantum fluctuations in the
couplings of the world-sheet field theory ($\sigma$ model)
characterizing the ``effective'' string theory sitting on
a typical classical string vacuum. In the
generic case, extra logarithmic operators appear, which
signal transitions between different conformal field-theory
backgrounds. Examples of this problem include the collective
coordinates of string solitons, among which are included
string black holes. The correct treatment of these
logarithmic operators is
essential for obtaining the appropriate finite
Zamolodchikov metric in theory space.
 
\pr
We have proceeded in section 3 to discuss the consistent
canonical quantization of string theory space. As we have
shown previously~\cite{emninfl}, the Liouville renormalization group approach
to this problem obeys the Helmholtz conditions which guarantee
the existence of such a canonical quantization scheme. We have
proposed a path-integral formulation (\ref{pi}) of this string field
theory, and demonstrated that it passes certain essential
consistency checks.
 
\pr
We went on to show in section 4 how this path-integral approach
may be used to treat extended-object backgrounds (solitons)
in string theory, demonstrating the emergence of the various
terms in the effective action (\ref{first}) as world-sheet
renormalization-group counterterms. In the recent string
literature, the collective coordinates characterizing such
solitonic backgrounds have been implemented in a simple and
elegant way by the imposition of suitable Dirichlet
boundary conditions on open-string world sheets. Duality
connects such $D$ branes with closed-string solitons. In
our path-integral approach (\ref{pi}), open strings and
$D$ branes emerge after an appropriate choice of the local
renormalization group scheme on the world sheet. The
precise formulation of string dualities within our
path-integral approach remains to be elucidated, but will
presumably impose interesting restrictions on the form of
the measure in the string path integral (\ref{pi}).
Using our Liouville renormalization
group approach, we have derived the appropriate
second-order equation of motion for the $D$ brane.
\pr
In section 5 we have applied this formalism to the discussion
of string black holes and the problem of information loss.
We have drawn the reader's attention to analogies between the
emerging higher-dimensional (HD) analysis of these questions
and our previous two-dimensional (2D) approach. In particular,
we have argued that the open strings appearing on the black-hole
horizon, treated as a $D$ brane in the HD approach, carry
Chan-Paton-like quantum numbers corresponding to the
$W_\infty$ charges first identified in the 2D model.
It is the leakage of information via these quantum numbers
which is responsible, within our interpretation, for
information loss from black holes.
 
\pr
Although we believe we have identified some key ingredients
in the eventual non-perturbative formulation of string field
theory, and shown how they may be applied to problems of
physical interest, we are amply aware that many important
formal issues remain to be elucidated. Among these, we
mention in particular the correct treatment of the logarithmic
operators associated with the back-reaction of matter on string
black-hole space times, which may be regarded as a string soliton
recoil problem. Also, as we have mentioned above, the important
restrictions imposed by duality on the measure of the string
path integral (\ref{pi}) remain to be explored. Moreover, we
have not considered in this paper the possible r\^ole of
supersymmetry, in particular in the stabilization of the
soliton backgrounds.
 
\pr
\nk {\Large {\bf Acknowledgements}}
\pr
We thank I. Kogan and J.F. Wheater for useful discussions.
The work of D.V. N. is partially supported by D.O.E. Grant
DEFG05-91-GR-40633.
\pr
\nk {\Large {\bf Note Added}}
 \pr
Since the appearance of our paper as hep-th/9605046,
two new papers have appeared which discuss the hair of $D$ branes and its
possible measurement \cite{StromD,BanksD}. The paper closer to ours in
its philosophy is \cite{BanksD}, which discusses the quantum hair of $D$
branes and the possible measurement of associated
generalized Aharonov-Bohm
phases in the scattering of highly-excited string states. These
ideas were
proposed previously in Refs. \cite{emnsel1,emnsel2}, on the basis
of explicit
studies \cite{earlieremn} in the $1\,+\,1$-dimensional black-hole model
of \cite{witt}. We linked these ideas to the appearance of
leg poles in
string scattering amplitudes, as also noted in \cite{BanksD}.
This author also mentions that BRST transformations become
non-trivial in the presence of $D$ branes: we made the same
remark in connection with world-sheet instantons related to
transitions among $1+1$-dimensional black holes in \cite{emn}.
In Sections 4 and 5
of this paper, we have developed the relation between our
earlier work and
subsequent $D$-brane studies, and shown how open string states carry
analogous $W$ quantum numbers.
\pr
The author of \cite{BanksD} states a belief that his
interpretation of his
results differs from ours \cite{emnsel1,emnsel2}. As
may be concluded from the previous paragraph, we agree
with the
above-mentioned aspects of his work, which is along the
lines of
\cite{emnsel1,emnsel2}. We have gone on elsewhere \cite{emn} to
develop the point of view that an
experimentalist capable only of low-energy experiments using lowest-level
string states would not be  able to disentangle all the string black-hole
states that could be distinguished by measurements using
highly-excited string states \cite{emnsel2,BanksD}.  The author
of ref. \cite{BanksD}
suggests that one may be able to reconstruct all the information
derivable from the scattering of excited string states
by considering external states with an arbitrarily large
number of lowest-level particles. It remains to be seen whether all the
black-hole information can be recovered this way in a
feasible experimental programme.
\pr
In the mean time, we observe that the
results of \cite{BanksD} confirm our
proposals in \cite{emnsel1,emnsel2}, and are in line with the results of
Section 5 of this paper.


\begin{thebibliography}{99}
\bibitem{emn} J. Ellis, N.E. Mavromatos
and D.V. Nanopoulos, Phys. Lett. B293 (1992), 37;
\par
{\it Lectures presented
at the
Erice Summer School, 31st Course: From Supersymmetry to the
Origin of Space-Time},
Ettore Majorana Centre, Erice, July 4-12
1993, published in Proc. Subnuclear Series Vol. 31, p.1
(World Scientific, Singapore 1994);
\par Mod. Phys. Lett. A10 (1995), 425;
and
hep-th/9305117.
\bibitem{zam} A.B. Zamolodchikov, JETP Lett. 43 (1986), 730;
Sov. J. Nucl. Phys. 46 (1987), 1090.
\bibitem{aben} I. Antoniadis, C. Bachas, J. Ellis
and D.V. Nanopoulos, Phys. Lett. B211 (1988), 393;
Nucl. Phys. B328 (1989), 117; Phys. Lett. B257 (1991), 278.
\bibitem{dvn} J. Polchinski, Nucl. Phys. 324 (1989), 123;
\par D.V. Nanopoulos, in {\it Proc. International
School of Astroparticle Physics}, HARC (Houston) (World
Scientific, Singapore, 1991), p. 183.
\bibitem{kogan} I. Kogan, preprint UBCTP-91-13 (1991);
Proc. {\it Particles and Fields 91}, p. 837, Vancouver
18-21 April 1991 (eds. D. Axen, D. Bryman and M.
Comyn, World Sci. 1992); see also Phys. Lett. B265 (1991), 269.
\bibitem{witt} E. Witten, Phys. Rev. D44 (1991), 314.
\bibitem{emninfl} J. Ellis, N.E. Mavromatos and D.V. Nanopoulos,
Mod. Phys. Lett. A10 (1995), 1685.
\bibitem{gibbons} A. Dabholkar and J. Harvey,
Phys. Rev. Lett. 63 (1989), 478;
\par A. Dabholkar, G. Gibbons, J. Harvey and
F. Ruiz Ruiz, Nucl. Phys. B340 (1990), 33;
\par for recent developments see A.A. Tseytlin,
Phys. Lett. B363 (1995), 223.
\bibitem{gur} V. Gurarie, Nucl. Phys. B410 (1993), 535;
\par M.A. Flohr, preprint CSIC-IMAFF-42, hep-th/9509166;
\par H.G. Kausch, preprint DAMTP-95-52, hep-th/9510149.
\bibitem{kogmav} I. Kogan and N.E. Mavromatos, preprint OUTP-95-50P,
hep-th/9512210, Phys. Lett. B in press.
\bibitem{horava} P. Horava, Phys. Lett. B289 (1992), 293.
\bibitem{callan} C. Callan and J. Maldacena,
preprint PUPT-1591, hep-th/9602043;
\par S. Das, preprint KEK-TH/472,TIFR-TH/96-11,
hep-th/9602172.
\bibitem{wormholes} S. Coleman, Nucl. Phys.
B310 (1988), 643;
\par T. Banks, I. Klebanov and L. Susskind, SLAC-PUB-4705
(1988).
\bibitem{DDK}F. David, Mod. Phys. Lett. A3 (1988), 1651;
\par J. Distler and H. Kawai, Nucl. Phys. B321 (1989), 509.
\bibitem{polchinski} J. Polchinski, Nucl. Phys. B307 (1988), 61;
{\it ibid.} B357 (1995), 241.
\bibitem{paban} W. Fischler, S. Paban and M. Rozali, Phys. Lett.
B352 (1995), 298.
\bibitem{schmid2} C. Schmidhuber,
Nucl. Phys. B435 (1995), 156.
\bibitem{aspects} J. Ellis, N.E. Mavromatos and D.V. Nanopoulos,
{\it Some Physical Aspects of Liouville String Dynamics},
contributions by J.E. and N.E.M. in {\it Phenomenology
of Unification from Present to Future}, Roma 23-26 March 1994,
p. 187 (World. Sci. 1994), hep-th/9405196.
\bibitem{bilal} A. Bilal and I. Kogan, preprint PUPT-1482,
hep-th/9407151 (unpublished); Nucl. Phys. B449 (1995), 569.
\bibitem{tsvelik} J.S. Caux, I.I. Kogan and A. Tsvelik,
preprint OUTP-95-62, hep-th/9511130.
\bibitem{conifold} For a concise recent review see:
A. Strominger, preprint UCSBTH-95-29, hep-th/9510207;
and references therein.
\bibitem{yung} A.V. Yung,
Int. J. Mod. Phys. A9 (1994), 591 ;
{\it ibid.} A10 (1995), 1553.
\bibitem{matrix} E. Br\'ezin and V.A. Kazakov,
Phys. Lett. B236 (1990), 144;
\par M. Douglas and A. Shenker, Nucl. Phys.
B335 (1990), 635;
\par D. Gross and A.A. Migdal, Phys. Rev. Lett.
64 (1990), 127;
\par For a
review see, e.g., I. Klebanov,
in {\it String Theory and Quantum Gravity}, Proc. Trieste
Spring School 1991, ed. by J. Harvey et al.
(World Scientific, Singapore, 1991), and references therein.
\bibitem{bhmatrix} A. Jevicki and T. Yoneya, Nucl. Phys.
B411 (1994), 64.
\bibitem{hojman} S.A. Hojman and L.C. Shepley, J. Math.
Phys. 32 (1991), 142 ;
\par F. Pardo, J. Math. Phys. 30 (1989), 2054.
\bibitem{osborn} H. Osborn, Nucl. Phys.
B294 (1987), 595; {\it ibid.}
B308 (1988), 629; Phys. Lett. B222 (1989), 97.
\bibitem{shore} G. Shore, Nucl. Phys. B286 (1987), 349.
\bibitem{review} For a review see A.A. Tseytlin, Int. J. Mod. Phys.
A4 (1989), 4249.
\bibitem{mavc} N.E. Mavromatos and J.L. Miramontes,
Phys. Lett. B212 (1988), 33;
\par N.E. Mavromatos, Phys. Rev. D39 (1989), 1659;
\par N.E. Mavromatos and J.L. Miramontes, Phys. Lett.
B226 (1991), 291.
\bibitem{ooguri} V. Periwal and  A. Strominger,
Phys. Lett. B235 (1990), 261;
\par S. Cecotti and C. Vafa, Nucl. Phys.
B367 (1991), 359;
\par H. Ooguri and C. Vafa, Nucl. Phys. B361 (1991), 469;
{\it ibid.} B367 (1991), 83.
\bibitem{duf} M. Duff, R. Khuri and X.J. Lu, Phys. Rep. 259 (1995), 213.
\bibitem{dbrane} J. Polchinski, Phys. Rev. D50 (1994), 6041  ;
NSF-ITP-95-122 preprint, hepth/9510017 ;
\par J. Polchinski, S. Chaudhuri and C. Johnson,
NSF-ITP-96-003 preprint, hepth/9602052,
and references therein.
\bibitem{shenker} S. Shenker, preprint Rutgers RU-95-53,
hep-th/9509132.
\bibitem{emn4d} J. Ellis, N.E. Mavromatos
and D.V. Nanopoulos, Phys. Lett. B278 (1992), 246.
\bibitem{bachas} C. Bachas, preprint NSF-ITP-95-144,
CPTH-S388-1195; hep-th/9511043.
\bibitem{tseytl2} A. A. Tseytlin, Phys. Lett. B363 (1995), 223.
\bibitem{mcavity} H. Osborn, Nucl. Phys. B363 (1991), 486;
\par D.
 McAvity and H. Osborn, Nucl. Phys. B406 (1993), 655;
Nucl. Phys. B455 (1995), 522.
\bibitem{dbrsc} C. Callan and I. Klebanov, preprint
PUPT-1578,
hep-th/9511173;
\par C. Schmidhuber, preprint PUPT-1585,
hep-th/9601003.
\bibitem{mnbrain} In a different context see: N. Mavromatos
and D.V. Nanopoulos, preprint
ACT-19/95, CTP-TAMU-55/95, OUTP-95-52P, quant-ph/9512021.
\bibitem{periwal} V. Periwal and O. Tafjord, preprint PUPT-1607,
hep-th/9603156.
\bibitem{rey} S.J. Rey, preprint SNUTP 96-030, hep-th/9604037.
\bibitem{liu} J. Polchinski, Nucl. Phys. B231 (1984), 413;
\par J. Hughes, J. Liu and J. Polchinski, Nucl. Phys.
B316 (1989), 15.
\bibitem{Mli} M. Li, preprint BROWN-HET-1027, hep-th/9512042.
\bibitem{hallc2} V. John, G. Jungman and S. Vaidya,
Nucl. Phys. B455 (1995), 505;
\par G. Amelino-Camelia and D. Bak, Phys. Lett. B343 (1995), 231.
\bibitem{yost} E.S. Fradkin and A.A. Tseytlin, Phys. Lett. B163 (1985), 123 ;
\par A. Abouelsaood,
C. Callan, C.R. Nappi and S.A. Yost, Nucl. Phys. B280 (1987) 599;
\par C. Callan, C. Lovelace, C.R. Nappi and S.A. Yost, Nucl. Phys. B308 (1988) 221.
\bibitem{tsc} C. Schmidhuber and
A.A. Tseytlin,
Nucl. Phys. B426 (1994), 187;
\par H. Dorn, preprint HU-Berlin-IEP-94-21, hep-th/9410084.
\bibitem{tsemem} A.A. Tseytlin, Phys. Lett. B251 (1990), 530.
\bibitem{kaln} S. Kalara and D.V. Nanopoulos, Phys. Lett.
B267 (1991), 343.
\bibitem{emnsel1} J. Ellis, N.E. Mavromatos and D.V. Nanopoulos,
Phys. Lett. B284 (1992), 27.
\bibitem{emnsel2} J. Ellis, N.E. Mavromatos and D.V. Nanopoulos,
Phys. Lett. B284 (1992), 43.
\bibitem{hallc} J. Ellis, N.E. Mavromatos and D.V. Nanopoulos,
Phys. Lett. B289 (1992), 25; {\it ibid.} B296 (1992), 40.
\bibitem{giveon} A. Giveon, Mod. Phys. Lett. A6 (1991), 2843.
\bibitem{wilc} F. Larsen and F. Wilczek, Princeton preprint, hep-th/9604134.
\bibitem{sussk} L. Susskind, Rutgers preprint,
hep-th/9309145;
\par L. Susskind and J. Uglum, Phys. Rev. D50
(1994), 2700.
\bibitem{chaudh} S. Chaudhuri and J. Lykken, Nucl. Phys B396 (1993),
270.
\bibitem{wgauge} S. Das, A. Dhar, G. Mandal and
S. Wadia, Int. J. Mod. Phys. A7 (1992), 5165.
\bibitem{bakas} I. Bakas and E. Kiritsis, Int. J. Mod. Phys. A7
(suppl.)  A7 (1992), 55; Phys. Lett. B301 (1993), 49.
\bibitem{wrapkog} In a rather different context,
the idea of wrapping the world-sheet
around the horizon of a spherically-symmetric
four-dimensional target-space black hole appeared first in:
I. Kogan, Mod. Phys. Lett. A6 (1991), 3297;
\par Subsequently a similar discussion appeared in
S. Giddings, J. Harvey, J. Polchinski, S. Shenker and
A. Strominger, Phys. Rev. D50 (1994), 6422.
\bibitem{mgreen} M. Green, Nucl. Phys. B293 (1987), 593.
\bibitem{ashworth}
I. Kogan, Phys. Lett. B256 (1991), 369;
Nucl. Phys. B375 (1992), 362;
\par S. Carlip, Nucl. Phys. B362 (1991), 111;
\par M.C. Ashworth, Mod. Phys. Lett. A10 (1995), 2749.
\bibitem{poly} J. Minahan and A. Polychronakos,
Phys. Lett. B312 (1993), 155.
\bibitem{das} S. Das and A. Jevicki, Mod. Phys. Lett.
A5 (1990), 1639.
\bibitem{lerda} R. Caracciolo, A. Lerda, and G. Zemba,
Phys. Lett. B352 (1995), 304.
\bibitem{pope} I. Bakas, Phys. Lett. B228 (1989), 57;
\par C. Pope, X. Shen and L. Romans, Nucl. Phys.
B339 (1990).
\bibitem{marcus} N. Marcus and A. Sagnotti,
Phys. Lett. B188 (1987), 58.
\bibitem{floratos} E. Floratos, J. Iliopoulos and
G. Tiktopoulos, Phys. Lett. B217 (1989), 285.
\bibitem{eguchi} T. Eguchi, Mod. Phys. Lett. A7 (1992), 85.
\bibitem{poly2} J. Minahan and A. Polychronakos,
Phys. Lett. B326 (1994), 288.
\bibitem{StromD} A. Strominger, {\it Statistical Hair
on Black Holes},
Rutgers
Univ. preprint RU-96-47,
hep-th/9606016.
\bibitem{BanksD} T. Banks, {\it Quantum
Hair on $D$-branes and Black-Hole Information in String
Theory},
Rutgers
Univ. preprint RU-96-49,
hept-th/9606026.
\bibitem{earlieremn} J. Ellis, N.E. Mavromatos
and D.V. Nanopoulos, Phys. Lett. B267 (1991), 465; {\it ibid.}
Phys. Lett. B272 (1991), 261.
\end{thebibliography}
\end{document}